\documentclass[preprintnumbers,amsmath,amssymb,prl]{revtex4-2}
\usepackage{graphicx}
\usepackage{dcolumn}
\usepackage{soul}
\usepackage[usenames,dvipsnames]{xcolor}
\usepackage{mathptmx}
\usepackage{amssymb}
\usepackage{array}
\usepackage{amsmath}
\usepackage{setspace}
\usepackage{hyperref}
\usepackage{comment}

\begin{document}
\preprint{ \color{Blue} \today}

\title{\Large \color{Blue} Grooves spacing govern water retention during condensation}

\author{M.Leonard, N.Vandewalle}
\affiliation{GRASP, Institute of Physics B5a, University of Li\`ege, B4000 Li\`ege, Belgium.}


\begin{abstract}
    Condensation on vertical surfaces leads to fluid retention, which limits the efficiency of applications ranging from heat exchangers to atmospheric water harvesters. A common strategy is to structure the surface with grooves, yet whether grooves help drainage or worsen retention remains unclear. Here we use a high-throughput condensation setup to quantify retention on substrates patterned with parallel vertical grooves of fixed geometry ($d/w=1$) while varying the spacing $s$. We uncover two opposite regimes separated by the droplet detachment radius $R_d$. For large spacings ($s>R_d$), droplets grow and slide under gravity while grooves, acting as passive reservoirs, increase retention. For small spacings ($s<R_d$), grooves instead trigger active drainage, confining droplet growth and reducing retention to values even lower than on smooth surfaces. Two asymptotic models, a groove-volume reservoir model and a plateau-packing model, capture this transition and explain the scaling of retention with $s$. These findings show that groove spacing controls whether grooves act as drains or reservoirs, providing a simple geometric design rule for tailoring condensation retention in practical systems.
\end{abstract}

\maketitle

\section{Introduction}

Dew drops on a leaf after a clear night are a familiar sight. But behind this everyday image lies a fascinating phenomenon. Some leaves, such as those of the lotus, prevent water from spreading across their surface. Instead, droplets remain nearly spherical and roll off easily, like marbles on wax. This so-called Lotus effect \cite{barthlott_purity_1997} is enabled by surface substructures that reduce adhesion between the liquid and the solid. More than a visual curiosity, this effect plays an essential biological role: it helps clean the surface and limits the accumulation of water that could promote disease \cite{lenz_ecological_2021}.

But what happens on a vertical surface when the drop is not deposited but forms through condensation? How do small-scale surface structures influence the retention and removal of condensed water under the pull of gravity? These questions matter not only in natural settings but also in engineering and technology. For instance, condensation can reduce thermal performance in heat exchangers such as air conditioners, refrigerators, or dehumidifiers. Droplets grow larger over time on smooth vertical surfaces and act as insulating barriers \cite{rose_dropwise_2002,kim_dropwise_2011}. A common approach would be to microtexture the surface to make it super hydrophobic, reducing droplet adhesion and allowing water to slide off. However, this strategy has limitations: water often condenses inside surface cavities and can remain trapped \cite{narhe_nucleation_2004,wier_condensation_2006}.

Similar challenges arise in atmospheric water harvesting. As global water scarcity becomes increasingly critical \cite{un_water_2020}, passive technologies such as fog collectors \cite{shi_fog_2018} and dew collectors \cite{muselli_dew_2009} are gaining attention. Dew collectors, in particular, typically yield only a few tens of centiliters of water per square meter each night. Because of this limited yield, even small retention losses matter. Unfortunately, from 15\% to 100\% of the condensed water remains stuck to the surface. Without active removal, this retained water tends to evaporate again during the day, wasting a scarce resource.

To enhance this natural removal process, researchers have turned to nature for inspiration. Surface structures such as grooves \cite{van_hulle_effect_2023,chen_ultrafast_2018,leonard_droplets_2023}, cones \cite{van_hulle_capillary_2021}, hairs \cite{andrews_three-dimensional_2011,protiere_wetting_2013}, and bumps \cite{park_condensation_2016} can guide and accelerate water transport at micro- and millimetre scales. Among these options, parallel vertical grooves stand out for their simplicity, efficiency, and compatibility with large-scale manufacturing. When closely packed (spacing $< 0.5\, \rm{mm}$), they reduce the latency time $\tau_L$, the delay before the first drop is collected under the substrate \cite{bintein_grooves_2019}, and maintain effective drainage over time, even under conditions of material ageing \cite{lavielle_plastic_2023}. They can also enhance the emissivity of the surface, increasing the overall condensation rate \cite{poualvarez_efficient_2025}. Yet their net effect on retention is unclear: do grooves act as drains that enhance water removal, or as reservoirs that increase storage?

This study investigates how the spacing $s$ between vertical grooves, ranging from $10.00\,\mathrm{mm}$ to $0.30\,\mathrm{mm}$, controls water retention during condensation for a fixed groove geometry. We find that spacing governs a transition between two competing drainage regimes: one driven by gravitational shedding, the other by groove-mediated transport. We capture this shift with two asymptotic models based on $s$. Beyond this transition, we explore how the groove aspect ratio $d/w$ modulates retention within the groove-driven regime. While we do not model geometry effects quantitatively, due to current limits in fabrication and theory, we uncover clear qualitative trends. Together, our results show that a single geometric variable can switch the role of grooves from passive reservoirs to active drains, offering a simple rule for designing surfaces that either hold or release water more effectively.

\section{\label{sec:level2}Experimental Setup}

\begin{figure}
    \centering
    \includegraphics[width=0.65\linewidth]{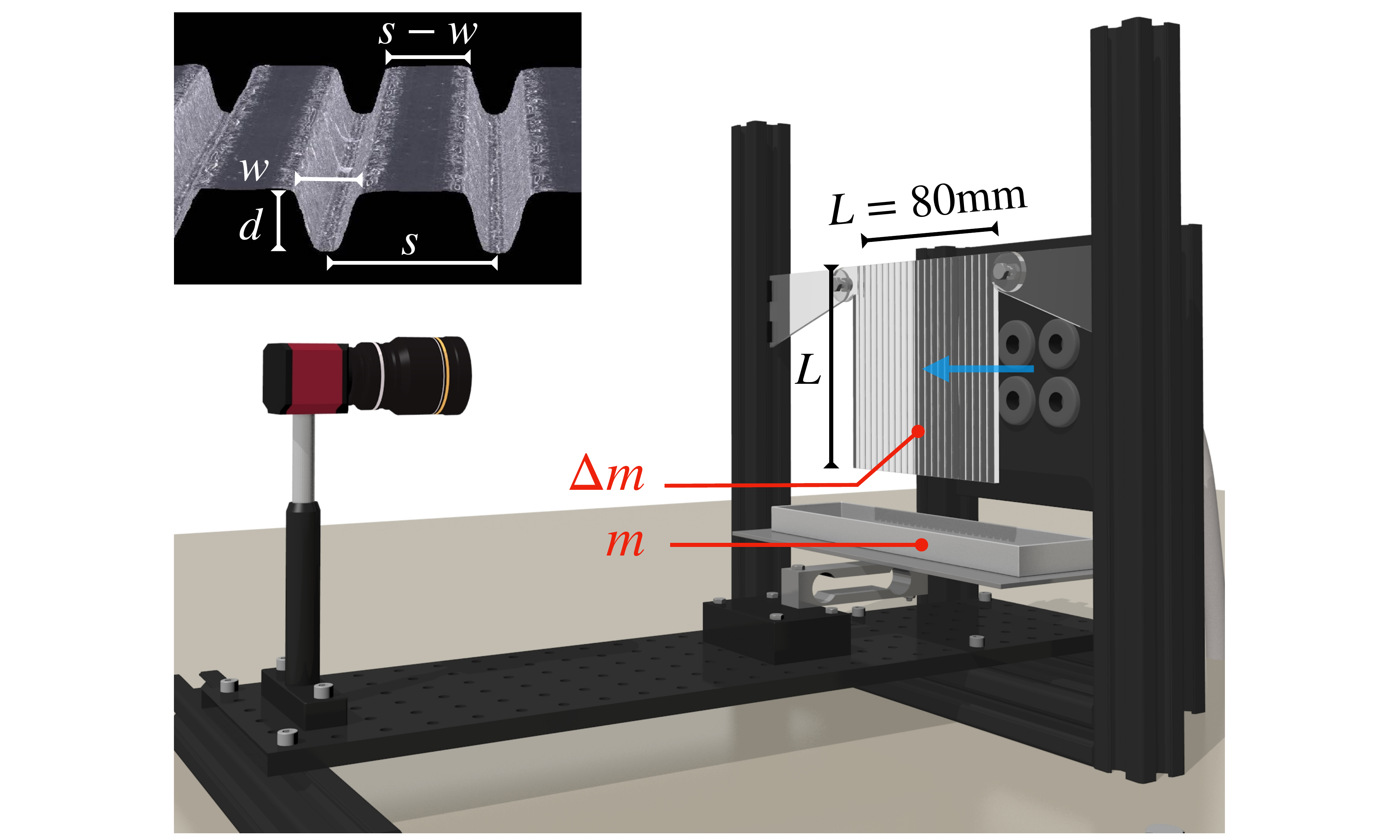}
    \caption{Experimental setup. Warm, humid air is generated by bubbling compressed air through a heated water reservoir and directed toward the vertically mounted substrate via four nozzles (right). Condensation forms on a square acrylic plate of side length $L = 80\,\mathrm{mm}$, either smooth ($s = 80.00\,\mathrm{mm}$) or patterned with parallel vertical grooves. The grooves have depth $d$ and width $w$ both equal to $0.20 \pm 0.01\,\mathrm{mm}$, while the spacing $s$ between grooves ranges from $0.30$ to $10.00\,\mathrm{mm}$. The flat region between two adjacent grooves is referred to as a plateau, with a width equal to $s-w$. Water detaching from the substrate is collected in a vessel connected to a strain gauge, providing the time-dependent collected mass $m(t)$. The retained mass $\Delta m(t)$, representing the water still present on the substrate, is obtained from $m(t)$ as described later in the text. The inset shows the groove cross-section, where the groove geometry parameters $d$, $w$, and $s$ are indicated.} 
    \label{fig:set_up}
\end{figure}

The traditional method of forced condensation involves cooling a surface by bringing it into contact with a thermal exchanger. Like condensation forming on a cold water bottle out of the fridge, water condenses when a surface is cooled below the dew point. While reliable, this technique is slow, typically yielding condensation rates around $6\,\rm{g/m^2\,h}$ \cite{trosseille_roughness-enhanced_2019,bintein_grooves_2019,jin_atmospheric_2017}. As a result, experiments often require several hours to complete.

Our approach takes the opposite route: instead of cooling the surface, we blow warm, humid air onto a room-temperature substrate, just like breathing on a cold window. Using this method, we reach rates of $900\,\rm{g/m^2\,h}$, 150 times faster, reducing experiment time and enabling many tests across a wide range of parameters. Experiments are performed in a climate-controlled chamber at $T = 20.0 \pm 0.5^{\circ}\rm{C}$ and relative humidity $RH = 65 \pm 2\%$. To generate the humid airflow, we heat a water reservoir to $75 \pm 2^{\circ}\rm{C}$. Compressed air is injected through a diffuser at the bottom of the reservoir. As it rises, the air warms up through thermal exchange and becomes saturated with water vapor. This warm, moist air exits through four nozzles located on the reservoir lid and flows perpendicularly toward the vertically suspended substrate at a velocity $v < 1\,\rm{m/s}$ (see Fig.~\ref{fig:set_up}). These velocities are comparable to those occurring in natural dew formation environments \cite{muselli_dew_2009,jacobs_passive_2008}.

The substrate reaches a temperature of $T = 45.0 \pm 1.5^{\circ}\rm{C}$ within about 100 seconds. At this temperature, liquid water has a density $\rho = 992\,\rm{kg/m^3}$ and surface tension $\sigma = 69 \times 10^{-3}\,\rm{N/m}$ \cite{gittens_variation_1969}, corresponding to a capillary length of $\lambda = \sqrt{\sigma / \rho g} = 2.67\,\rm{mm}$. The substrate is a square acrylic plate (TroGlass Clear) with thickness $3\,\rm{mm}$ and side length $L = 80\,\rm{mm}$. Its wetting properties are characterized by a static advancing contact angle $\theta_A = 78.0 \pm 4.5^\circ$ and a receding angle $\theta_R = 48.0 \pm 4.7^\circ$, yielding an average contact angle of $\theta = 63\pm 4.6^\circ$. Surface structuring is done with a laser cutter (Speedy 100, Trotec), producing grooves with spacing $s$ ranging from $0.30$ to $10.00\,\rm{mm}$, with depth $d$ and width $d$ such as $d=w=0.2 \pm 0.01\,\rm{mm}$. A smooth surface ($s = 80.00\,\rm{mm}$) serves as the reference. Groove depth $d$ and width $w$ are both measured using optical microscopy (Keyence VHX), see inset of Fig\ref{fig:set_up}. In the following, we refer to the flat regions between adjacent grooves, of width  equal to $s-w$, as plateaus.

During the experiment, condensed water flows downward under gravity and drips into a collection vessel. The collected water mass is measured in real-time with a strain gauge, with a resolution of $0.05\,\rm{g}$, approximately the mass of a single drop. Each experiment lasts 50 minutes and is repeated three times to ensure reproducibility. A CCD camera records images at one frame per second from behind the transparent substrate opposite the vapor source, as represented on Fig \ref{fig:set_up}.

\section{Transport description}

\begin{figure}
    \centering
    \includegraphics[width=1\linewidth]{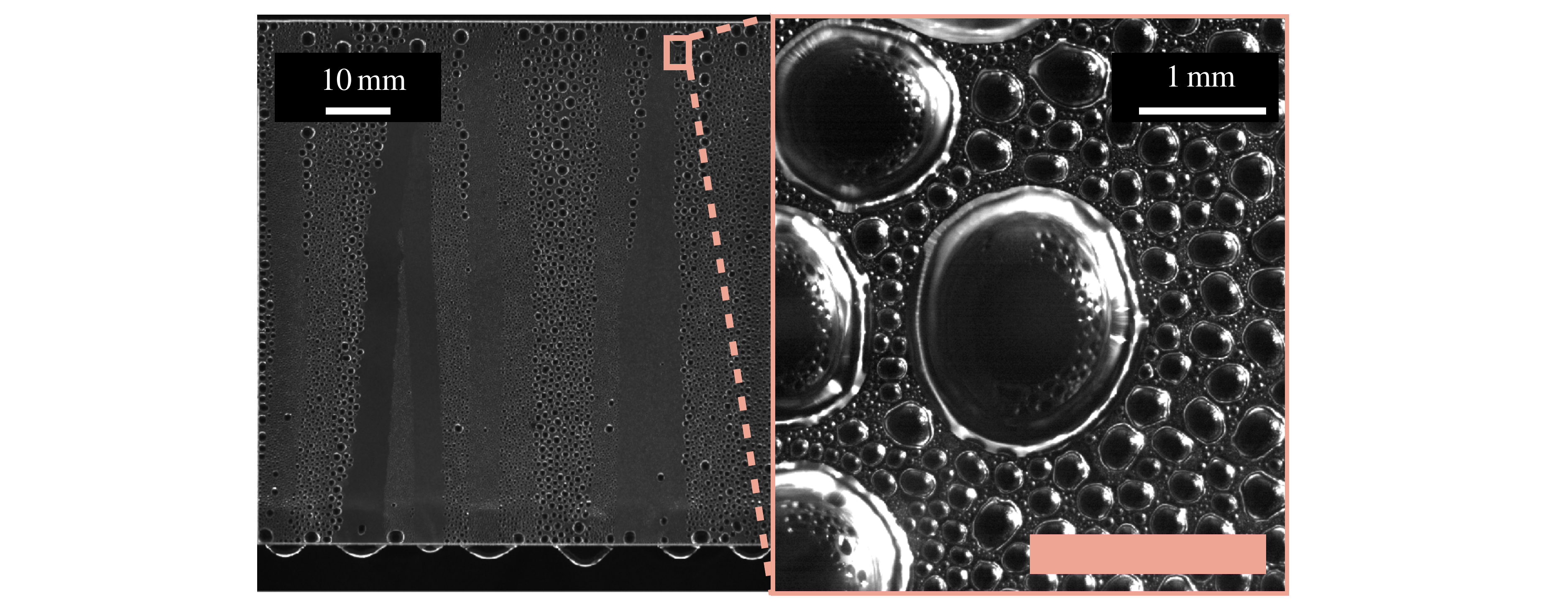}
    \caption{(Left) Overview of the smooth substrate, showing horizontal bands of similarly sized droplets. The droplet size varies from one band to another, reflecting the time elapsed since the last passage of a sweep droplet. These bands act as a short-term memory of the system: each is the trail left by a droplet sliding and sweeping smaller droplets from its path, with a characteristic trail width $2R_t$. The bands are not perfectly straight, as the sweep droplet’s trajectory can be deflected by coalescence with droplets encountered along the way. Larger droplets are found in the upper part of the sample, where they have more time to grow before being intercepted by others, making the upper region the primary origin of sweep droplets. Once formed, these droplets traverse the entire surface, providing efficient drainage exclusively through gravitational shedding. Full overview video available in supplementary material \textit{s80.avi}. (Right) Close-up view, showing droplets of various sizes. Growth is unconstrained by substrate geometry, leading to a broad size distribution prior to detachment.}
    \label{fig:pic_smooth}
\end{figure}

\begin{figure}
    \centering
    \includegraphics[width=1\linewidth]{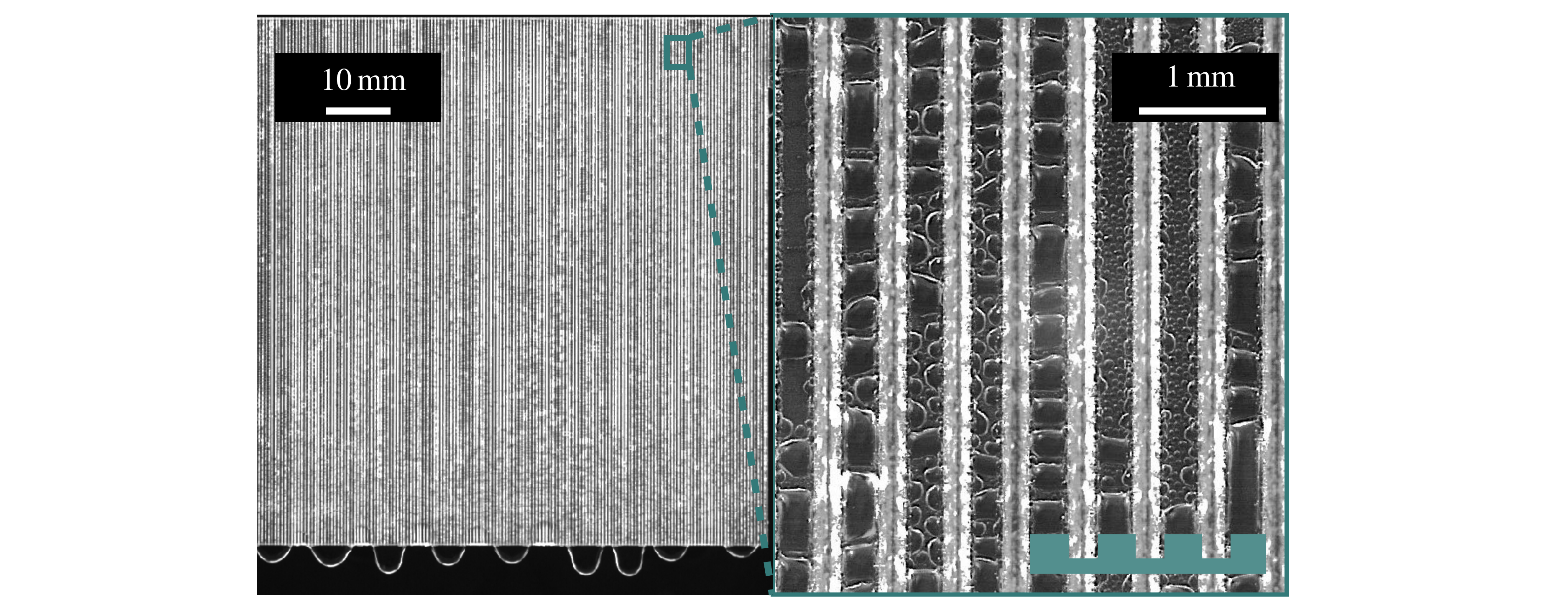}
    \caption{(Left) Global view of the grooved substrate ($s = 0.5\,\mathrm{mm}$, $d/w = 1$). In contrast to the smooth substrate, no large droplets are visible and no dry trails are present, indicating the suppression of gravitational shedding and the absence of drainage memory. Droplet size is limited by the plateau width, and transport occurs predominantly through groove-mediated drainage. Full overview video available in supplementary material \textit{s05dw1.avi}. (Right) Close-up view showing plateau and groove regions, as indicated by the schematic. Plateau droplets are laterally confined by the plateau width causing larger droplets to elongate along the groove direction.}
    \label{fig:pic_grooved}
\end{figure}

\subsection{Gravitational Shedding}

On a smooth vertical surface, condensation typically initiates at material imperfections, which act as preferential nucleation sites~\cite{lavielle_memory_2023}. Because the process starts uniformly and nearly simultaneously across the entire substrate, droplet nucleation and early growth are highly synchronized~\cite{bintein_grooves_2019}, a behavior also observed during natural dew formation and in engineered systems such as heat exchangers. Droplets grow initially by vapor adsorption and subsequently through coalescence with neighboring droplets. Once a droplet reaches critical radius $R_c$, its weight overcomes surface retention forces and it begins to slide downward~\cite{extrand_retention_1990,gao_how_2018}. As it descends, the droplet collects smaller droplets in its path~\cite{rose_dropwise_2002}, rapidly increasing its volume. Over time, its size stabilizes, through the emission of satellite droplets due to Rayleigh–Plateau instabilities~\cite{podgorski_corners_2001,le_grand_shape_2005}. This process leaves behind a narrow, nearly dry trail of width $2R_t \approx 2\lambda$, where $\lambda$ is the capillary length. The dry wake provides a clean region where new nucleation events can occur, effectively resetting the local condensation cycle. Occasionally, some of the satellite droplets released during descent are already close to the detachment radius and quickly begin to slide in turn. This creates a repeating cycle, adsorption, coalescence, sliding, instability, that governs condensation dynamics over time. Because droplets near the top of the surface experience less frequent sweeping, they have more time to grow and are more likely to initiate a shedding event. As a result, drainage is managed progressively, droplet by droplet, from the top part of the surface.

The fact that a sweep droplet collects all droplets in its path leads to an interesting property of the system. Once a region is scraped clean, it begins a new cycle of nucleation and growth, just like at the start of the experiment. Until it is scraped again, the droplets in that region grow at similar rates, resulting in a population of droplets with fairly uniform size. Moreover, in steady state, the formation of sweep droplets at the top of the sample is indepedant from one drop to an other. As a result, the system exhibits vertical bands of droplets whose sizes vary from one band to another, depending on the time elapsed since the last sweep droplet passage, as clearly visible on Fig. \ref{fig:pic_smooth} (Left). The striped pattern that forms across the sample thus represents the drainage memory of the system, a short-term memory, as each new band partially erases the previous one.

Finally, it is worth noting that droplets do not follow straight paths. Instead, their trajectories are influenced by interactions with the droplets they collect. During coalescence events, the fluid within the droplet reorganizes, which can alter its course. This effect is visible in Fig. \ref{fig:pic_smooth}(Left), where two trails that start very close to each other gradually diverge as they descend. This behavior, combined with the striped surface pattern, allows sweep droplets to deviate from a vertical trajectory and collect bands they would have only partially swept otherwise.

\subsection{Groove Drainage}
We now examine the grooved substrate with spacing $s = 0.5\,\mathrm{mm}$ and geometry $d = w = 0.2\,\mathrm{mm}$. Initially, nucleation resembles that on a smooth substrate. However, a novel behaviour emerges: droplet growth becomes strictly confined by the plateau width Fig.~\ref{fig:pic_grooved}. This geometric constraint prevents droplets from reaching the detachment radius, effectively suppressing classical gravitational shedding, an even extreme situation then described by Bintein \textit{et al} \cite{bintein_grooves_2019}, where droplets overspanning multiple grooves are always visible. A central question follows: how is water evacuated in this regime? Three physical mechanisms govern drainage under such confinement: the wetting interaction between a droplet and a groove, the drying process, and long-range coalescence (LRC). 

When a droplet interacts with a rectangular groove, its morphology depends on the contact angle $\theta$ and the aspect ratio $d/w$. According to Seemann \textit{et al.}~\cite{seemann_wetting_2005,seemann_wetting_2011}, three regimes arise. For small aspect ratios and large $\theta$, the droplet remains compact, spreading little into the groove (D regime). For moderate aspect ratios, the liquid fills the groove and forms a filament with positive Laplace pressure, the $F^+$ regime. This configuration is metastable for large volumes ($\Omega \gg w^3$), where it tends to revert to a droplet state. At high aspect ratios and low contact angles, the filament becomes fully confined with a concave meniscus and negative pressure, the $F^-$ regime. The transition to this stable groove-filling state is predicted by the following equation, linking the aspect ratio $d/w$ to the contact angle $\theta$
\begin{equation}
d/w = \frac{1-\cos{\theta}}{2 \cos{\theta}}.
\label{eq:seeman}
\end{equation}

In the $F^-$ regime, the groove acts as a one-dimensional capillary sink, drawing water from the adjacent plateaus. This process is fundamental to the drying phenomenon. It was first described by Narhe and Beysens~\cite{narhe_water_2006} during experiments on horizontal substrates with large-aspect-ratio rectangular grooves. As condensation continues, the water level within the groove rises until it contacts a droplet sitting on nearby plateaus. This initiates a rapid sequence of coalescence events: the initial merging raises the channel level, prompting further coalescence with neighbouring droplets. In a short time (20ms for a 1cm long groove in the original experiment), all droplets on the top surface of one or both grooves merge into the channel, leaving it dry. This process is driven entirely by surface tension, as droplets stay smaller than the capillary length $\lambda$. Once dry, the surface is immediately ready for new nucleation.

A final mechanism, long-range coalescence, enables fluid transfer between distant droplets connected by filled grooves. First reported by Bintein~\cite{bintein_grooves_2019}, LRC occurs over centimeter-scale distances through wetted grooves. Liquid flows from one droplet to another, driven by differences in Laplace or hydrostatic pressure. This mechanism allow the accelerate growth of droplets thus enabling earlier shedding. Water transport in vertically open groove, despite being the backbone of this mechanism, is not describe theoretically yet.

\begin{figure*}
    \centering
    \includegraphics[width=1\linewidth]{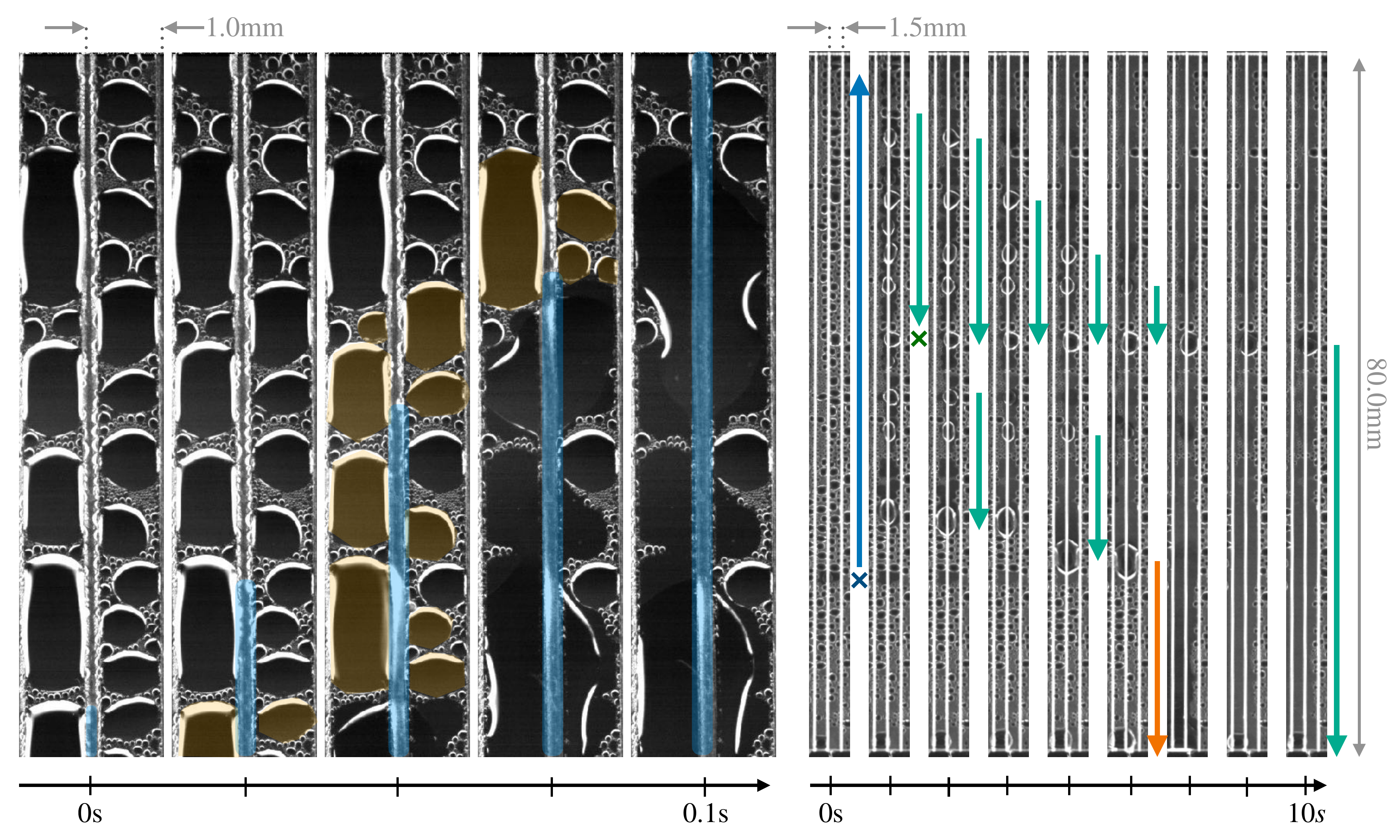}
    \caption{Drying dynamics and groove-mediated drainage. (Left) High-speed sequence on a substrate with $s = 1.0\,\mathrm{mm}$ ($d = w = 0.2\,\mathrm{mm}$) showing a capillary filament advancing upward inside a groove (blue), sequentially absorbing neighbouring plateau droplets (yellow). The process, independent of gravity, spreads at $\sim 100\,\mathrm{mm/s}$ and can leave overspanning droplets behind when coalescence exceeds groove capacity. (Right) Large-scale view of a drying event on $s = 1.5\,\mathrm{mm}$ spacing. The event initiates at the blue cross and propagates upward, redistributing water into the groove before sequential drainage begins (green arrows). Droplets are pumped in order of hydrostatic pressure, sometimes growing before being absorbed, illustrating hierarchical long-range coalescence. Near the origin, excess influx creates a large droplet that eventually sheds (orange arrow), highlighting the balance between groove drainage and gravitational shedding. Together, these dynamics show how drying funnels plateau water into grooves, with outcomes ranging from full absorption to overflow-induced droplet release.}
    \label{fig:drying}
\end{figure*}

To illustrate these concepts, we present in Fig.~\ref{fig:drying} (left) a drying event observed during our experiments on a substrate with spacing $s = 1.0\,\mathrm{mm}$ and groove geometry $d = w = 0.2\,\mathrm{mm}$. This configuration corresponds to an $F^-$ regime, as Equation~\ref{eq:seeman} gives a critical aspect ratio $d/w = 0.6<1$ for a contact angle $\theta = 63^\circ$. The sequence shows the time evolution ($\Delta t = 0.02\,\mathrm{s}$) of a defined region comprising a groove in the middle and two plateaus along. The advancing filament tip is highlighted in blue to aid visualisation. Initially, we see droplets sitting on both sides of the grooves. Then, as the filament goes upward (confirming that gravity plays no role in this phenomenon), droplets, highlighted in yellow, merge sequentially with the filament. This results in fast expansion of the filament at a rate of $100\rm{mm}/s$. The drying process can be partial as not all droplets sitting on plateaus were absorbed. Interestingly, we observe the formation of large overspanning droplets, resulting from the coalescence of plateau droplets.

To dig further into the implication of these straddling droplets during the drying process, we now look at what happens on the entire length of a groove on textured substrate ($s=1.5\rm{mm}$ and $d=w=0.2\rm{mm}$). This is represented in Fig. \ref{fig:drying} (right), where we pictured a groove at fixed time intervals ($\Delta t = 1.1s$). On the first frame, highlighted with a blue cross is the location where the drying process begins. Then, in the following picture, the influx propagated until the upper part of the sample, creating multiple droplets that span over the central groove. As explained earlier, this process is the result of a quick reorganisation of water distribution, from plateaus to the groove, and there is no drainage yet. In the following pictures, marked with a green arrows, we see gravitational drainage of the water through the groove. This drainage is sequential; higher droplets are pumped first, illustrating a specific case of long-range coalescence. Interestingly, we observe that the droplet at the mid-height of the sample, marked by a green arrow, appears to grow during the first 7 seconds and then shrinks, being pumped by the groove. This phenomenon emphasises the fact that there can be multiple hierarchies of pumping between the droplets, depending on Laplace and hydrostatic pressure, as mentioned earlier, as well as defect pinning or incomplete groove wetting, which likely alter the way fluid can circulate through the groove. 

An other interesting phenomenon is observed near the origin of the drying event, marked by a blue cross, where a large droplet undergoes a distinct two-stage evolution. In the first four frames, the droplet grows steadily, fed  through the groove by droplets located higher up. Then, in the fifth frame, it begins to slide down the surface, as visualised by an orange arrow. When the groove cannot drain the water fast enough, it leads to the formation of a droplet that may be large enough to shed. This highlights the interplay between groove-mediated drainage and gravitational shedding.

In summary, the interplay between coalescence, drying, and long-range coalescence gives rise to distinct water transport modes that depend strongly on substrate topography. On smooth substrates, drainage occurs exclusively via shedding: coalescence produces droplets large enough to overcome contact line pinning and slide under gravity, sweeping smaller neighbours along their path. In contrast, grooved substrates promote capillary confinement, shifting the dominant transport mechanism to groove drainage. Here, drying events funnel water from the plateaus into the grooves, where it flows downward under gravity. When this drainage is efficient, water remains fully confined, leaving no droplets visible on the surface. However, if the local influx exceeds groove capacity, overflow occurs. This can lead to short-lived straddling droplets (static overflow) or to larger droplets that shed under gravity (shedding overflow). Intermediate cases also emerge, where droplets begin to slide but are eventually reabsorbed by the groove, reflecting a dynamic balance between groove and gravitational drainage.

\section{Retention}

\subsection{Smooth substrate}

\begin{figure}
    \centering
    \includegraphics[width=0.7\linewidth]{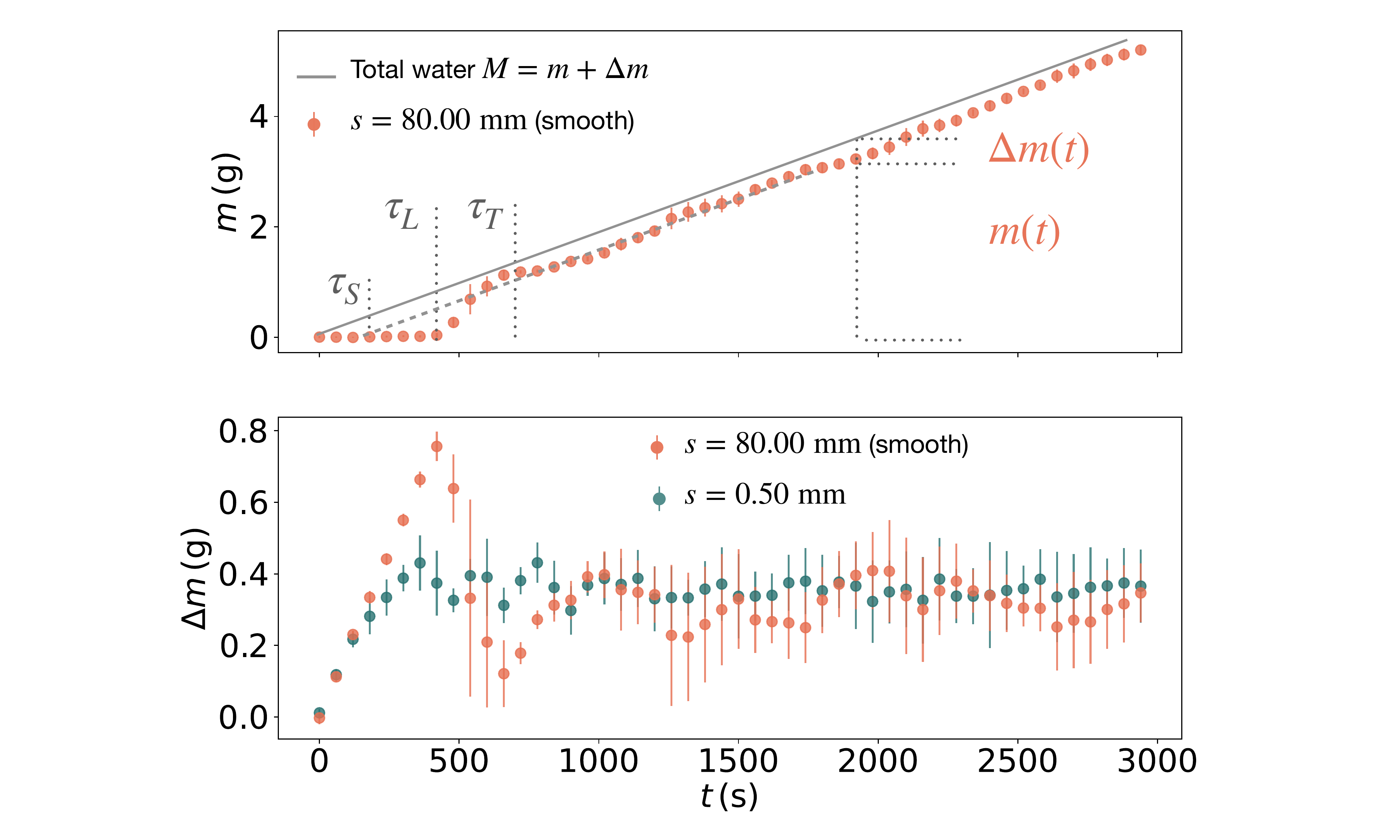}
    \caption{Condensation dynamics on smooth and grooved substrates. (Top) Collected water mass $m(t)$ on a smooth surface shows three regimes: a latency period with no collection ($0\rightarrow\tau_L$), a transient burst when synchronized droplets reach the detachment radius $R_d$ ($\tau_L\rightarrow\tau_T$), and a steady state where $m(t)$ grows linearly at a constant condensation rate $c$ ($\tau_T\rightarrow\infty$). The grey line indicates the total condensed mass $M(t)$, with the vertical gap to $m(t)$ giving the retained mass $\Delta m(t)$. The extrapolated time $\tau_S$ marks when the first droplet would arrive if the system were operating directly in steady state. (Bottom) Retained mass $\Delta m(t)$ for different groove spacings $s$. Narrow grooves shorten the latency and suppress the transient burst, but the steady-state retention $\Delta m$ remains comparable to the smooth case. This challenges the view that shorter latency always $\tau_L$ signals higher drainage efficiency, showing instead that $\Delta m$ is the key metric for long-term performance.}
    \label{fig:m_dm}
\end{figure}
\subsubsection{Results}
To establish a baseline, we first examine condensation and drainage on a smooth vertical substrate. Fig.~\ref{fig:m_dm}(top) shows the collected water mass $m(t)$ over time, while Fig.~\ref{fig:m_dm}(bottom) presents the retained mass $\Delta m(t)$. Both curves reveal three distinct regimes. In the latency regime ($0\rightarrow\tau_L$), $m(t)$ remains at zero, indicating that no droplet has yet reached the collection vessel, while $\Delta m(t)$ increases steadily as droplets grow on the surface. The transient regime ($\tau_L\rightarrow\tau_T$) begins once the first droplet arrives at the vessel, producing a sharp rise in $m(t)$ and a corresponding drop in $\Delta m(t)$ as retained water is evacuated. Finally, in the steady-state regime ($\tau_T\rightarrow\infty$), $m(t)$ increases linearly with a constant slope, and $\Delta m(t)$ fluctuates around a stable average. Let's now dive further in each of these regime.

The latency regime corresponds to the period when condensation is already active but all water remains on the surface. Droplets grow until they reach the detachment radius $R_d$, the smallest size at which they slide under gravity. Classical theory predicts motion at a critical radius
\begin{equation}
    R_c = \lambda \sqrt{\frac{\cos\theta_R - \cos\theta_A}{A}} = 1.80\,\mathrm{mm},
\end{equation}
with $A = 0.33$ related to the drop cap geometry~\cite{beysens_dew_2018}. However, in our experiments, droplet motion frequently occurs at radii smaller than the theoretical critical radius $R_c$. We define the observed onset of sliding as the detachment radius, denoted $R_d$
\begin{equation}
    R_c > R_d > 1\,\mathrm{mm}.
\end{equation}
We attribute discrepancy between $R_c$ and $R_d$ to the dynamic nature of coalescence during condensation. When two droplets merge, the resulting droplet undergoes a rapid reconfiguration, during which the triple contact line moves significantly. Gao \textit{et al.}~\cite{gao_how_2018} showed that the friction opposing droplet motion is considerably lower when the droplet is already in motion than when it is static, an effect analogous to solid-on-solid friction. This reduction in resistance during contact line displacement likely facilitates the onset of sliding at smaller droplet sizes, thereby explaining the difference between the theoretical critical radius $R_c$ and the experimentally observed detachment radius $R_d$. As they slide down the surface, these droplets sweep smaller droplets along their path, leaving behind a nearly dry trail. These descending droplets accumulate on the lower edge of the substrate, where they remain hanging until they detach. The moment the first droplet detaches and reaches the collection container defines the latency time $\tau_L$.

Let us now turn to the burst of collection that marks the transient regime. Its origin lies in the initial synchronicity of droplet growth, which keeps the droplet size distribution narrow at early times. As a result, many droplets reach the detachment radius $R_d$ almost simultaneously and slide down within a short interval, producing a sharp, temporary increase in the collected water mass $m(t)$ just after $\tau_L$. This synchrony also allows us to estimate the theoretical maximum retention of the smooth substrate during latency. Assuming droplets of radius $R_d = 1\,\mathrm{mm}$ arranged in a hexagonal lattice, the configuration that maximises surface occupancy, the packing fraction is $\epsilon = 0.91$, representing the ratio of droplet-covered to total surface area. The volume of a single droplet is given by~\cite{beysens_dew_2018}
\begin{equation}
    \Omega_d = A\pi R_d^3,
\end{equation}
with $A$ a geometrical factor linked to droplet shape. The corresponding maximum retained mass is
\begin{equation}
    \Delta m_{L}^{\mathrm{max}} = \frac{\epsilon L^2}{S_d} \rho \Omega_d = 1.92\,\mathrm{g},
\end{equation}
where $S_d$ is the droplet contact surface. This value is more than twice the experimental peak of $\Delta m$ observed in Fig.~\ref{fig:m_dm}(bottom), showing that in practice, imperfect synchrony and coalescence-induced shedding prevent the surface from reaching this upper bound. The estimate therefore represents a worst-case retention scenario, assuming perfect packing and no early drainage, and will vary with substrate wetting properties through $A$. After this initial burst, synchrony progressively decays, marking the transient regime ($\tau_L \rightarrow \tau_T$), until the system approaches steady state.

Finally, the steady-state regime commences when the collection rate stabilises, defining a straight slope in $m(t)$ graph of Fig.~\ref{fig:m_dm}(top). Thermodynamic constraints govern this rate: for condensation to occur, the latent heat of vaporisation must be removed. In our setup, heat is dissipated by thermal exchange with ambient air maintained at $20^\circ\mathrm{C}$ in the climate chamber. Because this process does not depend on the surface texture~\cite{jin_atmospheric_2017,trosseille_roughness-enhanced_2019}, the condensation rate remains constant across all substrates, with a measured value of $c = 1.9 \; 10^{-3}\,\mathrm{g/s}$. It allows us to estimate the total amount of water condensed $M(t)$ at any time $t$. Indeed, it corresponds to a straight line of slope $c$ passing through the origin of the graph. As we measure the collected water $m(t)$ and we know the total amount of water condensed $M(t)$, we deduce the amount of water retained on the surface $\Delta m(t)$ during the steady state phase, such as 
\begin{equation}
    \Delta m(t) = M(t) - m(t)
\end{equation}
In Fig.~\ref{fig:m_dm}(top), these three quantities are represented. The grey line with slope $c$ is the total amount of water condensed $M(t)$, dots represent the water collected in the vessel $m(t)$. The vertical gap between this reference and the experimental $m(t)$ curve provides a dynamic estimate of $\Delta m(t)$. Once the system reaches the steady state ($t > \tau_T$), the retained mass $\Delta m$ fluctuates around a stable average value, as visible on Fig. \ref{fig:m_dm}(bottom).

\subsubsection{Model}
Can we predict the retained mass on a smooth substrate, $\Delta m_s$, during the steady-state regime? We propose a semi-empirical model that considers only the water resting directly on the surface, explicitly excluding hanging droplets, which will also be neglected in the following models. These suspended droplets (visible on the bottom part of Figs \ref{fig:pic_smooth} $\&$ \ref{fig:pic_grooved}) are modeled as half-ellipsoids, with the major axis corresponding to the droplet width, the minor axis to the substrate thickness ($3\,\rm{mm}$), and the height taken from image measurements. Their volume, measured through image analysis, accounts for about 5–10\% of the total retained mass $\Delta m$.

We measured the time interval $\Delta t$ between successive droplet sliding events, along with the average width of the trails they leave behind. From these measurements, we estimate a trail width $2R_t = 5.34 \pm 0.88\,\mathrm{mm}$. The mean time interval between events is $\Delta t = 22.13 \pm 21.85\,\mathrm{s}$. The fact that the standard deviation is of the same order as the mean suggests that droplet shedding follows a Poisson process, in which events occur independently at a constant average rate. In such a regime, the waiting times between events are exponentially distributed, reflecting the absence of temporal correlation between successive sliding events. Based on this, we estimate a characteristic surface renewal time
\begin{equation}
t_r = \frac{\Delta t \, L}{2 R_t} = 331.5 \,\mathrm{s}.
\end{equation}
This time is also the characteristic time of drainage memory. It obviously depends on the condensation rate $c$. From this, we infer the average retained mass on a smooth substrate
\begin{equation}\label{eq:gravity}
\Delta m_s = \frac{c \, t_r}{2} = 0.31 \,\mathrm{g},
\end{equation}
where $c$ is the measured condensation rate. This prediction agrees well with experimental measurements ($\Delta m_s = 0.35 \pm 0.08\,\mathrm{g}$), although it remains a simplified model that neglects the complex dynamics of coalescence-driven growth.

In summary, for a smooth substrate, the retained water mass is mainly determined by the time it takes for droplets at the top to grow, detach, and slide down, sweeping up others along the way. This time, and thus the average retention, depends on the condensation rate and both detachment radius $R_d$ and trail width $2R_t$. When scaled to a $1\,\rm{m^2}$ surface, the retained mass reaches about $54\,\rm{g}$ of water, or roughly $5\,\rm{cL}$. This volume corresponds to nearly 15\% of the total dew that can be collected on a productive night \cite{muselli_dew_2009,jacobs_passive_2008}, and highlights the importance of minimizing retention to improve the efficiency of passive water collectors.

\subsection{Spacing Variation}

\subsubsection{Results}
To ensure stable groove drainage conditions, we carefully set their aspect ratio in accordance with the condition outlined in Equation~\ref{eq:seeman}. Specifically, with an average contact angle of $\theta = 63^\circ$, this leads to the requirement that $d/w > 0.6$. Therefore, we select $d/w = 1.00$ with $d=w=0.20 \pm 0.01\rm{mm}$ to accommodate potential imperfections in the substrate.

We start by examining the amount of water retained on a sample with a groove spacing of $s = 0.50\,\mathrm{mm}$, as shown in Fig.~\ref{fig:m_dm}(bottom). Several observations emerge. First, the latency time $\tau_L$ is shorter. Second, the intermediate regime is nearly absent. Finally, despite these two aspects suggesting that this sample may be more efficient than the smooth reference, the actual amount of water retained on the surface during the steady state regime, $\Delta m$, is found to be quite similar. This observation challenges the previously proposed link between latency time and surface efficiency~\cite{bintein_grooves_2019}.

To capture this effect, we introduce a new characteristic time, $\tau_S$. It estimates how long it would take for the first droplet to be collected if the system started directly in steady state. We obtain $\tau_S$ by extrapolating the steady-state $m(t)$ curve back to the time axis (Fig.~\ref{fig:m_dm}, top):
\begin{equation}
\tau_S(s) = \frac{\Delta m(s)}{c},
\label{eq:tau_S}
\end{equation}
where $\Delta m(s)$ is the steady-state retained mass for spacing $s$, and $c$ is the condensation rate.

\begin{figure}
    \centering
    \includegraphics[width=0.9\linewidth]{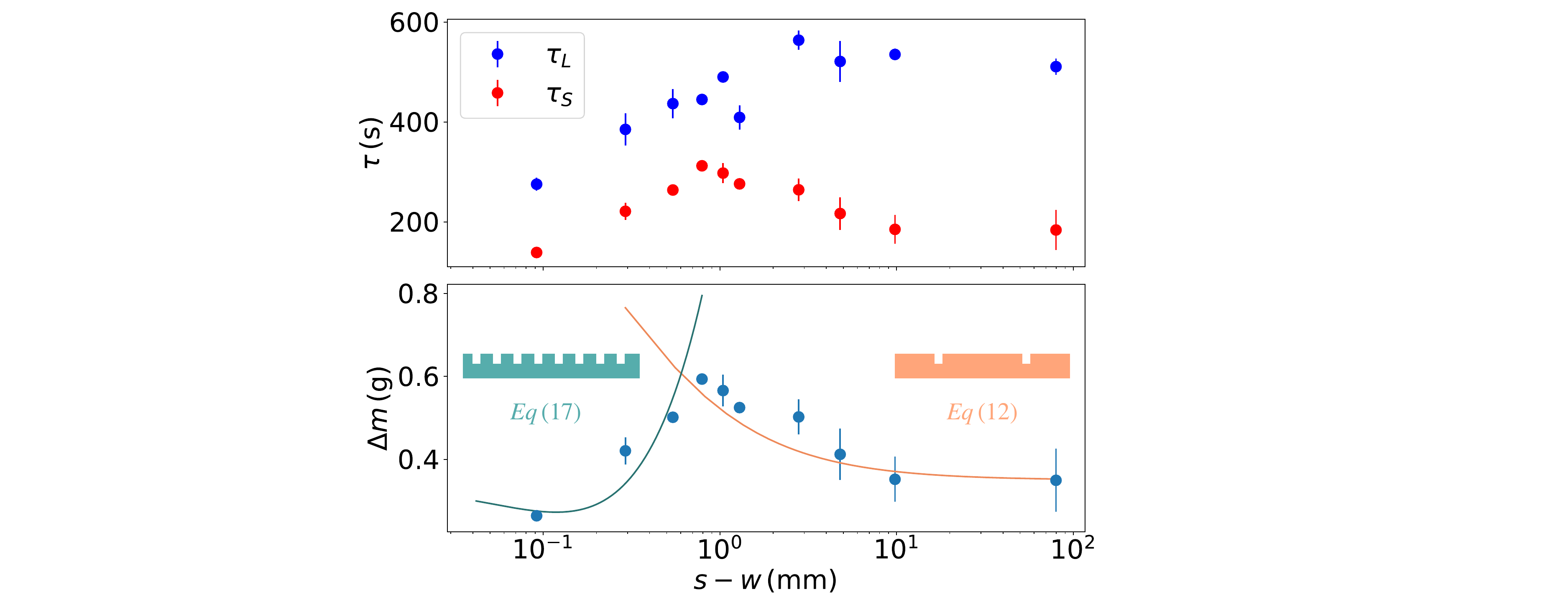}
    \caption{(Top) Comparison of latency time $\tau_L$ (blue) and extrapolated steady-state time $\tau_S$ (red) as functions of plateau width $s-w$, with $w=0.2\,\rm{mm}$ held constant. For small $s$, both times scale similarly, indicating negligible transient effects. For larger $s$, the increasing divergence between $\tau_L$ and $\tau_S$ reflects excess water retention from early-stage synchronized droplet growth. (Bottom) Total retained water mass $\Delta m$ as a function of plateau width $s-w$. Experimental data (blue circles) reveal two opposing trends: retention increases with $s$ for densely grooved substrates ($s < 1\,\mathrm{mm}$), then decreases for larger spacings, approaching the smooth-surface limit. Orange and green curves correspond to asymptotic models valid for sparsely (Eq. \ref{eq:sparsly}) and densely grooved (Eq. \ref{eq:densely}) regimes, respectively.}
    \label{fig:data_spacing}
\end{figure}

Fig.~\ref{fig:data_spacing}(top) compares $\tau_S$ with the measured latency time $\tau_L$. For small spacings ($s < R_d$), the two times scale proportionally, showing that densely grooved substrates exhibit little or no transient regime. Beyond $R_d$, however, the curves diverge: $\tau_L$ saturates near $s \approx 3\,\mathrm{mm}$, while $\tau_S$ decreases. This divergence reflects a transient accumulation of water caused by synchronized droplet growth. The excess is progressively drained as the system relaxes into steady state. In practice, substrates with a $\tau_L$ similar to the smooth surface all display the same burst of retained mass, revealing that $\tau_L$ predicts the onset of collective shedding but not the long-term efficiency of drainage, as clearly visible on Fig.~\ref{fig:m_dm}(bottom).

This observation leads us to compare samples based on the average retained mass $\Delta m$ during the steady-state regime. These values are reported in Fig.~\ref{fig:data_spacing}(bottom) for groove spacings ranging from $s = 0.30\,\mathrm{mm}$ to $s = 80.00\,\mathrm{mm}$, the latter serving as a smooth-surface reference. In this figure, we express the horizontal axis not in terms of $s$, as done previously, but in terms of the plateau width $s - w$. Since the groove width $w$ is held constant at $0.2\,\mathrm{mm}$, the plateau width is directly proportional to $s$, and the underlying control parameter remains the groove spacing. As expected from Equation~\ref{eq:tau_S}, $\Delta m$ exhibits a similar trend to that of $\tau_S$. Two opposing regimes can be identified. In the range $s - w < 1.00\,\mathrm{mm}$, the retained mass $\Delta m$ increases with increasing plateau width. However, beyond $1.00\,\mathrm{mm}$ and up to the smooth-surface case, this trend reverses: further increasing the plateau width leads to a decrease in the amount of retained water.

\begin{figure*}
    \centering
    \includegraphics[width=0.9\linewidth]{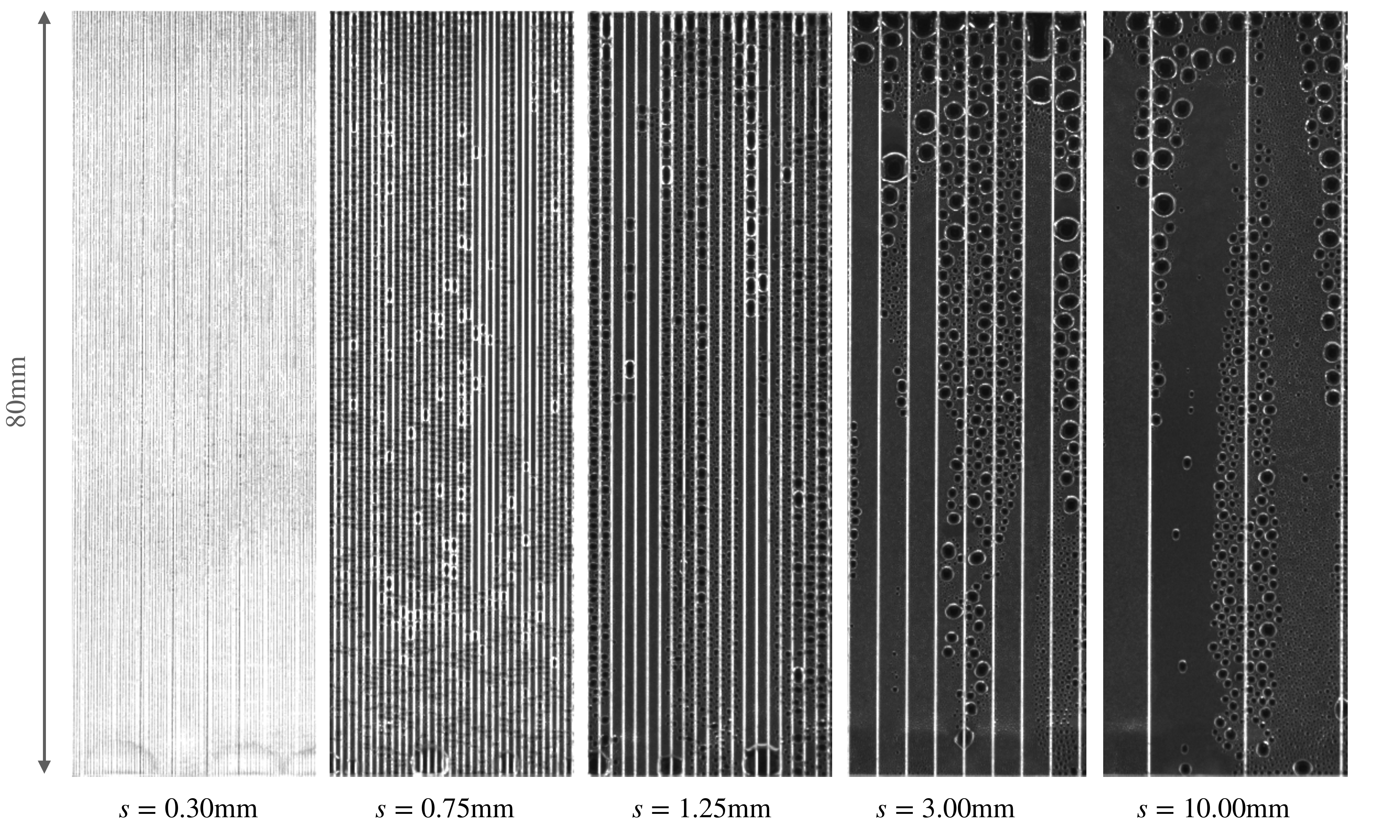}
    \caption{Snapshots of condensation patterns in the steady state ($t > 2000\,\mathrm{s}$) for substrates with groove spacing increasing from left to right: $s = 0.30$, $0.75$, $1.25$, $3.00$, and $10.00\,\mathrm{mm}$. At large spacing ($s \approx 10\,\mathrm{mm}$), the surface resembles a smooth substrate, with wide striped bands formed by sweep droplets sliding under gravity and leaving trails of width $2R_t$, occasionally accompanied by satellite drops. At intermediate spacing ($s \approx 3\,\mathrm{mm}$), the bands narrow and droplet size becomes limited by $s$, constraining vertical coalescence. Drying events become more frequent, generating straddling droplets that can evolve into sweep droplets mid-height, although these often stop part-way and are absorbed by grooves, suppressing efficient drainage. At very small spacing ($s \approx 0.75\,\mathrm{mm}$), sweep droplets vanish altogether. Straddling droplets formed during drying remain pinned and are reabsorbed, leaving only laterally confined droplets with no correlation in size across the surface. The progression highlights a transition from gravity-driven drainage with memory effects (right) to groove-constrained dynamics dominated by groove drainage (left). 
    }
    \label{fig:deltaM-s}
\end{figure*}

To understand this antagonistic behaviour, we analyse images taken in the steady state ($t > 2000\,\mathrm{s}$) for various spacings ($s=0.30-0.75-1.25-3.00-10.00\,\rm{mm}$) represented in Fig. \ref{fig:deltaM-s}. For large spacings ($s \approx 10\,\mathrm{mm}$) the surface resembles that of a smooth substrate. One observes the characteristic striped pattern made up of droplet clusters with similar sizes. Coalescence events produce sweep droplets, which then slide freely, clearing a trail of width $2R_t$. Occasionally, satellite droplets are emitted along the way. Since droplets are not geometrically confined by the grooves, successive coalescence events during growth tend to shift their center of mass away from the groove axis. As a result, apart from a few isolated drying events, primarily observed during the early stages of condensation, the grooves appear to contribute little to drainage compared to gravitational shedding.

When the spacing is reduced ($s \approx 3\,\mathrm{mm}$) the striped pattern becomes tighter, with band widths now limited by the spacing $s$. As a result, grooves limit the size of static droplets within each band, constraining coalescence along the vertical direction and limiting the spontaneous formation of sweep droplets. At the opposite, drying events are more frequent, generating straddling droplet along the groove. The lowest droplet generated is typically the one that has the lowest hydrostatic pressure. It drains fluid from upward droplets and evolves into a straddling sweep droplet. Thus, sweep droplets are no longer initiated exclusively from the top of the sample, but can also form mid-height, as illustrated in Fig.~\ref{fig:drying}. As $s$ decreases, sweep droplets are less likely to reach the bottom of the sample. Instead, they stop mid-way and are absorbed by a groove, further reducing drainage efficiency. Finally, these sweep droplets no longer leave satellite droplets behind. Since they straddle a groove as they descend, any potential satellite droplet is rapidly absorbed by the groove.

In the limit of very narrow spacing ($s \approx 0.75\,\mathrm{mm}$), even sweep droplets generated by drying are absent. Occasionally, groove spanning droplets appear during drying episodes, but they remain fixed and are eventually absorbed by the groove. This results in a droplet distribution strictly limited in width by $s$, highly dispersed on the sample and lacking any clear size correlation between neighbouring droplets. There is nothing left similar to the smooth case.

\subsubsection{Model}

The control variable in our model is the plateau width, $s-w$, since it defines the space where droplets can grow and coalesce. In our experiments, however, the groove width $w$ was held fixed. This means that, in practice, all variations in $s-w$ come from changing $s$ alone. With this in mind, we develop two asymptotic models that capture the dominant transport modes in the limits of large and small spacing. These models are not intended to describe every detail of the fluid dynamics, as fundamental theory is lacking on several key concepts (drop morphology on a grooved vertical substrate, coalescence driven shedding, open groove drainage, drying phenomenon, ...)  but rather to provide physical insight into how retention evolves as groove spacing is varied.

In the large-spacing limit ($s > R_d$), the plateaus between grooves are wide enough to permit unconstrained droplet growth and lateral coalescence. In this regime, gravitational shedding is the dominant drainage mechanism. The grooves, while occasionally activated, act mainly as passive reservoirs and play only a secondary role in transport.

We model retention in this regime by assuming it is equivalent to that of a smooth substrate, $\Delta m_s$, but scaled by the surface fraction occupied by plateaus:
\begin{equation}
\Delta m_{\mathrm{plateau}} = \frac{\Delta m_s}{L^2} n_g (s-w) L = \Delta m_s \frac{s-w}{s},
\end{equation}
where $n_g = L/s$ is the number of grooves on the substrate, $L$ is the substrate length, and $w$ is the groove width. The grooves themselves contribute an additional retained volume given by
\begin{equation}
\Delta m_{\mathrm{grooves}} = n_g \rho L d w = L^2 \rho \frac{d w}{s},
\end{equation}
where $d$ is the groove depth and $\rho$ is the fluid density. The total retained mass for a sparsely grooved substrate is thus
\begin{equation}
    \Delta m = \Delta m_s \frac{s-w}{s} + L^2 \rho \frac{d w}{s},
    \label{eq:sparsly}
\end{equation}
shown as the orange curve in Fig.~\ref{fig:data_spacing}(bottom).

This formulation assumes that the grooves remain mostly filled on average, which is consistent with observations that gravitational shedding dominates drainage when spacing is large. While this approach simplifies groove dynamics, it is expected to hold in situations where wide plateaus allow for unconstrained droplet growth and only limited interaction with grooves. We acknowledge that grooves can still be intermittently activated, occasionally altering droplet trajectories or triggering overflow events, but these effects are not explicitly captured in the model. They are assumed to have a secondary influence on overall retention, as supported by the good agreement between the model predictions and experimental measurements in this regime.

In the opposite limit of narrow groove spacing ($s < R_d$), the drainage mechanism changes fundamentally: grooves are frequently activated and handle most of the water removal. As before, we model the total retention as the sum of contributions from grooves and plateaus.

We begin with the groove contribution. Grooves act as narrow reservoirs that can trap a residual amount of water even after drainage events, thanks to continuous condensation for example. To capture this effect, we assume that the volume stored in grooves scales with the reference retention $\Delta m_s$, but weighted by the fraction of the surface occupied by grooves:
\begin{equation}
\Delta m_{\mathrm{grooves}} = \frac{\Delta m_s}{L^2} n_g wL = \Delta m_s \frac{w}{s},
\end{equation}
where $n_g = L/s$ is the number of grooves across the sample. This simple form reproduces the two natural limits. When $s \to w$, the plateaus shrink to nothing, and the surface is essentially all grooves. In that case, retention tends to the smooth reference value, $\Delta m_{\mathrm{grooves}} \to \Delta m_s$. At the opposite extreme, when $s$ becomes large, the number of grooves drops, and their contribution to storage vanishes, $\Delta m_{\mathrm{grooves}} \to 0$.

We now turn to the plateaus. These regions host droplets that are laterally confined by the plateau width $s-w$. To approximate their contribution, we assume that the median droplet diameter scales with the plateau width. This gives a droplet volume
\begin{equation}
\Omega_d = \frac{A \pi}{8}(s-w)^3,
\end{equation}
where $A$ is a geometrical factor linked to droplet shape. Because droplets are confined by plateau edges and often appear elongated (as in Fig.~\ref{fig:pic_grooved}), we take the packing fraction to be unity. This assumption also accounts for the presence of smaller droplets that fit between larger ones. The average number of droplets per plateau is then
\begin{equation}
n_d = \frac{4L}{\pi (s-w)}.
\end{equation}
Multiplying by the number of plateaus $n_p = L/s$, the total mass stored on plateaus becomes
\begin{equation}
\Delta m_{\mathrm{plateau}} = n_p n_d \Omega_d \rho = \frac{3}{8}L^2 \rho A \frac{(s-w)^2}{s}.
\end{equation}

Here too, the asymptotic behaviours are clear. When $s \ll w$, plateaus vanish, and their contribution goes to zero. As $s$ increases, droplet size scales with the growing plateau width, so $\Delta m_{\mathrm{plateau}}$ rises sharply. However, this growth cannot continue indefinitely. Once $s$ exceeds the detachment radius $R_d$, droplets on plateaus begin to shed under gravity, and retention will be capped by shedding dynamics rather than geometry. For this reason, the model should not be extrapolated beyond $R_d$ without additional corrections.

Combining both contributions, the total retained mass for densely grooved substrates is
\begin{equation}
\Delta m = \Delta m_s \frac{w}{s} + \frac{3}{8}L^2 \rho A \frac{(s-w)^2}{s}.
\label{eq:densely}
\end{equation}
Plotted as the green curve in Fig.~\ref{fig:data_spacing}(bottom), this expression captures the observed scaling of retention with groove spacing and agrees well with the experimental data in the dense-groove regime. It highlights how retention increases with groove density and how plateau-confined droplets dominate storage when $s < R_d$.

The model, however, is deliberately simple. It neglects processes likely active in this regime. Stable filaments may channel long-range capillary transfer or act as temporary reservoirs. Overflow shedding, when drying-induced droplets grow large enough to sweep the surface, is also ignored. Furthermore, the formulation depends only on spacing $s$, without explicitly accounting for groove geometry ($d$, $w$); that role is explored separately in the next section. Despite these omissions, the model provides a useful framework: it clarifies why retention scales with $s$ and explains the shift in control from grooves to plateau-confined droplets.

Interestingly, the lowest retention is found in this densely grooved regime, where groove-mediated drainage dominates and droplet growth is strongly confined. However, the exact location of this minimum, and its alignment with the model prediction, remains uncertain. This is due to fabrication limits: in our study, groove patterns were engraved using a CO$_2$ laser cutter, whose resolution prevents precise control of the plateau width below $0.1\,\mathrm{mm}$. To test the model’s asymptotic predictions more robustly, and to resolve the minimum more accurately, finer manufacturing methods will be required. Photolithography, for example, could enable a more systematic exploration of extreme groove densities.

At the junction of the two models, near $s - w = 0.61\,\mathrm{mm}$ (i.e., $s = 0.81\,\mathrm{mm}$), the predicted maximum droplet radius is $(s-w)/2 \approx R_d/4$. This value appears much smaller than the detachment radius $R_d$. However, this gap can largely be attributed to the fact that drying events often produce straddling droplets, which span across two plateaus. These droplets reach a characteristic radius of $(2s + w)/2$, which for $s = 0.81\,\mathrm{mm}$ gives $0.9\,\mathrm{mm}$, already close to $R_d$.

Why does a small difference remain? Part of the answer lies in the limitations of the model itself. First, $R_d$ is an experimentally observed threshold rather than a sharply defined value, and its precise determination would require further analysis. Second, groove confinement forces droplets into elongated shapes (Fig.~\ref{fig:pic_grooved}), making the spherical approximation inadequate. The groove edges may also act as pinning sites, increasing the apparent base angle and allowing confined droplets to hold more volume than spherical ones. Finally, the model does not capture the dynamics of droplets on grooved, and especially wetted, substrates, which may alter detachment thresholds.

Together, these effects highlight the complexity of the intermediate regime. Here, transport no longer follows a single dominant pathway: sweep droplets often originate mid-height during drying, their trajectories shaped by overflow and local wetting, and they may either slide or be reabsorbed. Neither asymptotic model fully captures this hybrid state, where coalescence, drying, and groove-mediated drainage overlap.

\subsection{Groove aspect ratio variation }
\begin{figure*}
    \centering
    \includegraphics[width=0.9\linewidth]{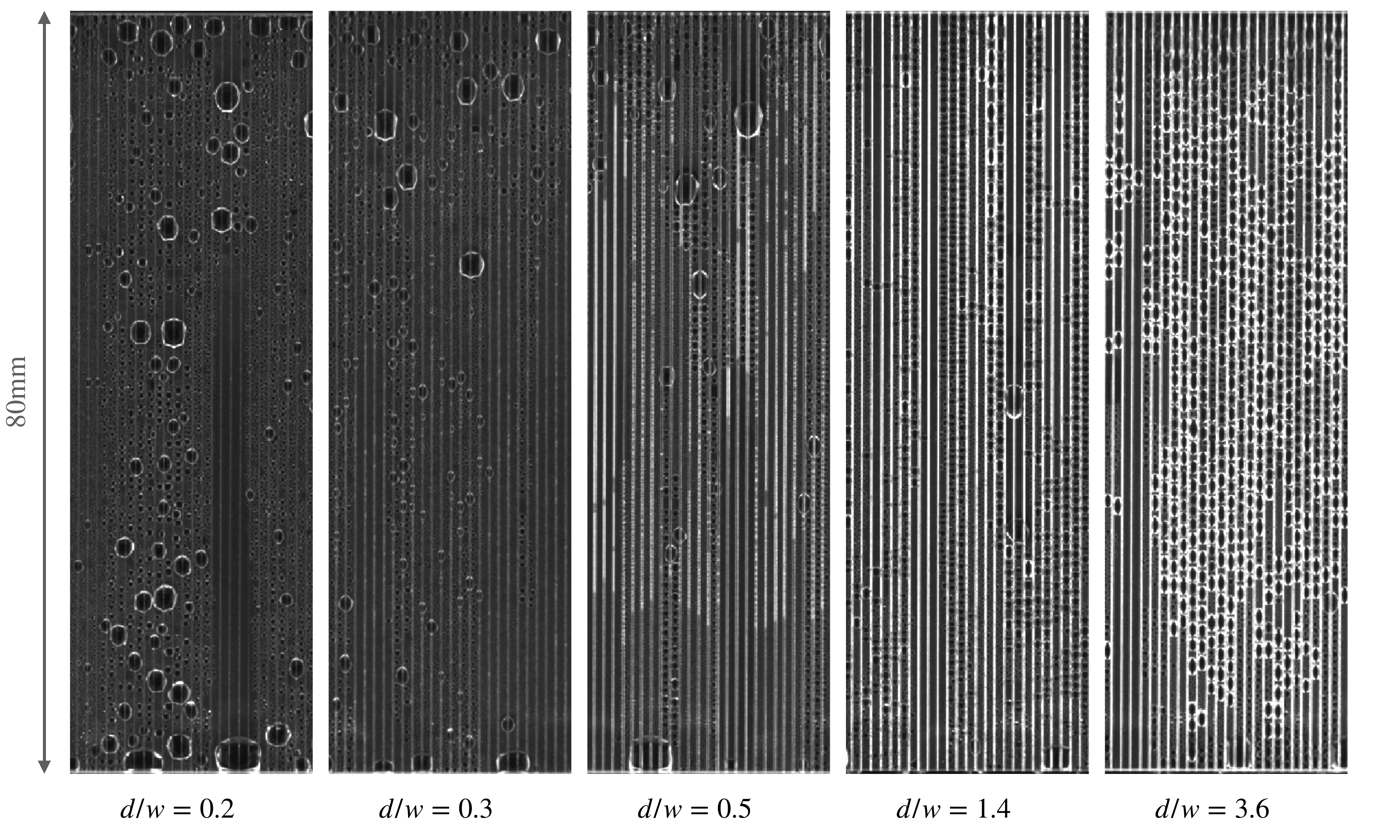}
    \caption{Snapshots of condensation patterns on grooved substrates with fixed spacing $s = 1.0\,\mathrm{mm}$ and increasing groove aspect ratios, from $d/w = 0.2$ to $3.6$. As $d/w$ increases, both droplet morphology and drainage mechanisms evolve. For shallow grooves ($d/w = 0.2$–$0.5$), droplet growth is still dominated by adsorption, coalescence, and gravitational shedding, but long-range coalescence (LRC) begins to play a role, producing large straddling droplets during the latency phase and gradually eroding drainage memory. Sweep droplets emerge lower on the substrate and with smaller radii as $d/w$ increases. At larger aspect ratios ($d/w \geq 1$), lateral coalescence is strongly suppressed, and transport shifts to groove-mediated drainage. Droplets become geometrically confined to plateaus or grooves, while overflow-driven straddling droplets are only occasionally observed and typically reabsorbed. This progression highlights a transition from shedding-dominated to groove-dominated drainage as groove depth increases.}
    \label{fig-aspect-ratio}
\end{figure*}

In the previous section, we showed that fixing the groove geometry in the $F^-$ regime, favorable to capillary imbibition, and varying the spacing $s$ leads to distinct drainage mechanisms, each associated with different retention behaviors. One underlying assumption in the densely grooved model was that groove geometry does not significantly affect retention. To test the robustness of this assumption, and to probe the limits of groove-controlled drainage, we now explore the role of groove aspect ratio $d/w$ under drainage-dominated conditions ($s \leq R_d = 1\,\mathrm{mm}$). This section is deliberately exploratory: our aim is not to develop a full model, but to identify qualitative transitions in droplet morphology and transport dynamics as $d/w$ increases.

Figure~\ref{fig-aspect-ratio} presents a series of substrates with fixed spacing $s = 1\,\mathrm{mm}$ and increasing aspect ratios from $d/w = 0.2$ to $3.6$. Let us first describe the visible changes induced by this variation. On a smooth substrate ($d/w = 0$), droplets grow through adsorption and coalescence, with no directional constraints. The early phase of condensation is marked by simultaneous and spatially uniform nucleation, leading to global synchronization. This collective growth triggers a burst of shedding during the intermediate regime. Transport is driven entirely by gravitational shedding: droplets typically detach near the top of the substrate, sweep smaller droplets along their path, and leave behind a clean trail of width approximately $2\lambda$. This trail resets nucleation locally, resulting in bands of similarly sized droplets, an expression of short-term drainage memory. The droplet population spans a broad range of sizes.

As the groove aspect ratio increases ($d/w = 0.2$ to $0.5$), the system enters a more complex regime. Droplet growth still occurs via adsorption and coalescence, but long-range coalescence begins to play a significant role. Coalescence becomes increasingly constrained in the vertical direction as grooves deepen. LRC, combined with classical coalescence, leads to the early emergence of large straddling droplets during the latency phase. This process introduces local desynchronization of droplet sizes, which progressively erodes the drainage memory. Transport remains dominated by shedding but is now assisted by LRC that enhances growth speed. Because LRC promotes the growth of downstream droplets, sweep droplets tend to initiate lower on the substrate. The radius of sweeping droplets also decreases with increasing $d/w$, narrowing from 4–5mm for $d/w = 0.2$ to 2–3mm ($\approx 2s$) at $d/w = 0.5$. The static droplet population becomes geometrically constrained as $d/w$ increases. Plateau and groove droplets are clearly distinct, shaped by the substrate’s topography. Straddling droplets appear more frequently, triggered by drying events, and their maximum diameter decreases with $d/w$, from $2\lambda$ down to approximately $2s$. These droplets can be stable or transient, momentarily involved in a drying phenomenon.

For large aspect ratios ($d/w \geq 1$), droplet growth occurs primarily through adsorption and vertical coalescence, with lateral interactions largely suppressed by the groove geometry. Transport is dominated by groove drainage, which becomes the main pathway for water removal. Occasionally, we observe overflow events, characterised by temporary straddling droplets appearing above the groove. These droplets are either reabsorbed shortly afterward or, in rarer cases, grow large enough to become sweep droplets. However, the formation of such mobile droplets becomes increasingly rare as the aspect ratio increases. At high $d/w$, groove absorption is more localized and less frequent: overflow events tend to affect only short segments of the substrate. The static droplet population is clearly split between plateau and groove droplets, each constrained in shape and size by the underlying texture.

\begin{figure}
    \centering
    \includegraphics[width=0.9\linewidth]{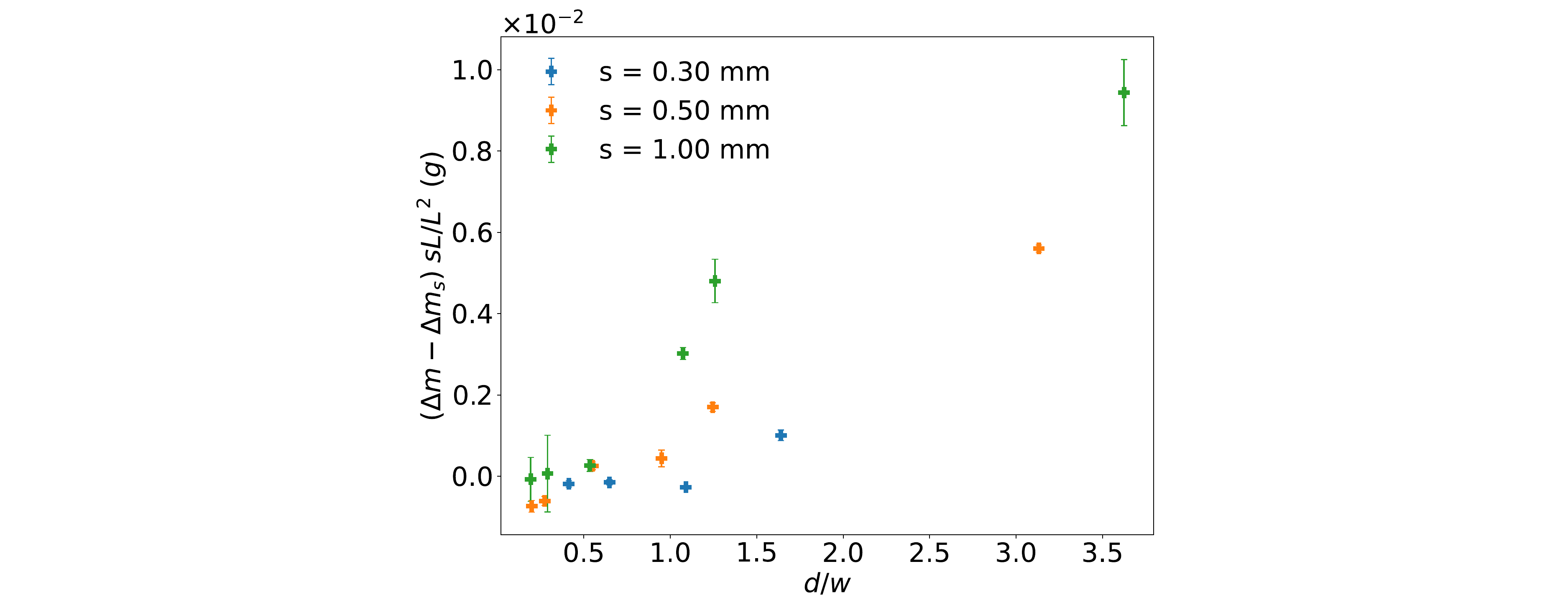}
    \caption{Normalized excess retention per plateau unit, $(\Delta m - \Delta m_s) \cdot s / L$, as a function of aspect ratio $d/w$, for three groove spacings ($s = 0.3$, $0.5$, and $1.0\,\mathrm{mm}$). For small $d/w$, retention is comparable to that of smooth substrates and largely independent of groove geometry. For larger $d/w$, retention increases with groove depth, indicating a transition to geometry-controlled storage. The position of this transition depends on the spacing $s$, with smaller $s$ delaying the onset of geometry sensitivity.}
    \label{fig:data_depth}
\end{figure}

We analyze how groove geometry influences water retention by varying the aspect ratio $d/w$ from 0.2 to 3.5, across three groove spacings representative of the densely grooved regime: $s = 1.0$, $0.5$ and $0.3\,\mathrm{mm}$. To isolate the role of geometry, we compare retention not in absolute terms, but by normalizing the excess mass relative to a smooth surface and scaling it to the plateau area. Specifically, we plot $(\Delta m - \Delta m_s) \cdot sL / L^2$, which quantifies how efficiently a groove–plateau unit evacuates water compared to an unstructured surface. Results are shown in Fig.~\ref{fig:data_depth}.

Two distinct regimes emerge. For low aspect ratios ($d/w \leq 1$), the curves remain flat and close to zero, indicating that groove geometry has little impact on retention. The differences observed between the $s$-dependent curves confirm that, in this regime, spacing governs drainage. This supports the assumption made in the densely grooved model: when $s \ll R_d$ and $d/w \approx 1$, the groove geometry can be neglected in first approximation.

As $d/w$ increases further, retention begins to rise, especially for $s = 1.0\,\mathrm{mm} \approx R_d$. This marks the onset of a second regime where groove geometry does influence retention. In this case, the filled-groove model provides a better description, as more water accumulates within the groove volume. The transition from the flat to the sloped regime depends on spacing: the smaller the spacing $s$, the later this shift occurs. This behavior reflects the fact that, for a droplet to interact with the groove, it must grow large enough to span the plateau. Narrower spacing reduces this required size, increasing the frequency of groove–droplet interactions and enhancing drainage efficiency. As groove depth increases, so does its storage capacity, contributing to higher retention per groove–plateau unit.

However, care must be taken when interpreting these trends. Since retention is expressed per unit plateau, the total mass stored on a sample also depends on the number of grooves, which increases as spacing decreases.

Comparing Fig.s~\ref{fig:deltaM-s}(left) and \ref{fig-aspect-ratio}(left) reveals consistent patterns. Substrates with dense grooves ($s = 0.75\,\mathrm{mm}$, $d/w = 1$) and those with deeper grooves at slightly wider spacing ($s = 1.0\,\mathrm{mm}$, $d/w = 3.6$) both limit droplet size and channel transport through the grooves, suppressing gravitational shedding. The difference lies in where water is stored: higher $d/w$ increases groove capacity, while larger $s$ shifts storage toward the plateaus.

In contrast, wide-spacing, moderate-depth substrates ($s = 10.0\,\mathrm{mm}$, $d/w = 1$) and tight-spacing, shallow-grooved ones ($s = 1.0\,\mathrm{mm}$, $d/w = 0.2$) both sustain large, unconstrained droplets and gravity-driven drainage. Yet the pathways differ. Wide spacing promotes sweep droplet formation through classical coalescence, largely independent of the grooves. Shallow grooves encourage long-range coalescence (LRC), which accelerates local droplet growth and promotes faster shedding events.

Together, these comparisons highlight a central principle: drainage and retention are governed by a balance of fluxes, between how much water condenses and how efficiently it can be evacuated. In densely grooved substrates, where gravitational shedding may be suppressed, this balance becomes especially critical. The input flux is primarily set by the available condensing area, determined by the plateau width $s - w$, while the evacuation flux depends on the groove geometry, particularly the aspect ratio $d/w$. In this study, we focused mainly on how groove spacing modulates the input side of this balance by altering the condensing surface. A full picture of drainage dynamics, however, will require exploring the output flux: how groove shape and internal wetting behavior control the capacity and rate of water evacuation. Capturing this interplay remains a key challenge for future work.

\section{Summary and outlook}

This study demonstrates that vertical grooves can fundamentally alter water retention during condensation, shifting the dominant drainage mode from gravitational shedding to groove drainage. By systematically varying groove spacing $s$ with a fixed aspect ratio $d/w=1$, we identified two limiting regimes. In the sparse-groove regime ($s > R_d$), retention is governed by gravitational shedding, with grooves acting primarily as passive reservoirs. In the dense-groove regime ($s < R_d$), droplets are confined to the plateaus and grooves are frequently activated, making groove-mediated drainage the primary pathway for water removal. Between these limits lies a transition zone, near the detachment radius $R_d$, where coalescence, drying, and groove transport overlap to produce hybrid drainage modes. Remarkably, water retention peaks in this intermediate region, illustrating the competition between two transport modes.

We developed two asymptotic models, each capturing the scaling of retained mass in one of the limits, and showed that their divergence near $R_d$ marks the boundary between the two dominant drainage mechanisms. Our results also reveal a previously unreported geometry-constrained condensation regime, in which droplets never exceed the plateau width. This confinement suppresses gravitational shedding and promotes groove-driven transport. Such surfaces could offer tailored thermal and optical properties, relevant for both passive water harvesting and functional heat-transfer surfaces.

Realising this potential will require higher-resolution fabrication to access smaller $s$ values, the development of a theoretical description for drainage within vertically open grooves, and a systematic exploration of how groove geometry (e.g., $d/w$) influences drainage within the groove-dominated regime. In this study, we modeled the dominant role of spacing $s$ in governing the drainage regime, while the influence of groove geometry was explored qualitatively due to fabrication and theoretical limitations. Future work could also investigate combined geometric metrics, such as $dw/s$, to quantify how both confinement and groove capacity interact. Together, these advances would provide a framework for designing textured surfaces with optimised drainage and retention characteristics, enabling more efficient control of condensation-driven processes.

\section{Supplementary Material}
Full overview videos from Fig.~\ref{fig:pic_smooth} (s80.avi) and Fig.~\ref{fig:pic_grooved} (s05dw1.avi). Complete videos of Fig \ref{fig:drying} are also available (left s10dw1.avi - right s15dw1.avi).

\section{Acknowledgments}
This work was financially supported by Valsem Industries SAS. The authors would like to thank Prof. Beysens for fruitful discussions, Prof. Parmentier for the microscope measurements, Mr. Melard and Mr. Rondia for their contributions to the experimental setup, and Mr. Corbisier for the experimental diagram

\section*{Data Availability Statement}
The data that support the findings of this study are available from the corresponding author upon reasonable request.


\begin{thebibliography}{32}%
\makeatletter
\providecommand \@ifxundefined [1]{%
 \@ifx{#1\undefined}
}%
\providecommand \@ifnum [1]{%
 \ifnum #1\expandafter \@firstoftwo
 \else \expandafter \@secondoftwo
 \fi
}%
\providecommand \@ifx [1]{%
 \ifx #1\expandafter \@firstoftwo
 \else \expandafter \@secondoftwo
 \fi
}%
\providecommand \natexlab [1]{#1}%
\providecommand \enquote  [1]{``#1''}%
\providecommand \bibnamefont  [1]{#1}%
\providecommand \bibfnamefont [1]{#1}%
\providecommand \citenamefont [1]{#1}%
\providecommand \href@noop [0]{\@secondoftwo}%
\providecommand \href [0]{\begingroup \@sanitize@url \@href}%
\providecommand \@href[1]{\@@startlink{#1}\@@href}%
\providecommand \@@href[1]{\endgroup#1\@@endlink}%
\providecommand \@sanitize@url [0]{\catcode `\\12\catcode `\$12\catcode `\&12\catcode `\#12\catcode `\^12\catcode `\_12\catcode `\%12\relax}%
\providecommand \@@startlink[1]{}%
\providecommand \@@endlink[0]{}%
\providecommand \url  [0]{\begingroup\@sanitize@url \@url }%
\providecommand \@url [1]{\endgroup\@href {#1}{\urlprefix }}%
\providecommand \urlprefix  [0]{URL }%
\providecommand \Eprint [0]{\href }%
\providecommand \doibase [0]{https://doi.org/}%
\providecommand \selectlanguage [0]{\@gobble}%
\providecommand \bibinfo  [0]{\@secondoftwo}%
\providecommand \bibfield  [0]{\@secondoftwo}%
\providecommand \translation [1]{[#1]}%
\providecommand \BibitemOpen [0]{}%
\providecommand \bibitemStop [0]{}%
\providecommand \bibitemNoStop [0]{.\EOS\space}%
\providecommand \EOS [0]{\spacefactor3000\relax}%
\providecommand \BibitemShut  [1]{\csname bibitem#1\endcsname}%
\let\auto@bib@innerbib\@empty
\bibitem [{\citenamefont {Barthlott}\ and\ \citenamefont {Neinhuis}(1997)}]{barthlott_purity_1997}%
  \BibitemOpen
  \bibfield  {author} {\bibinfo {author} {\bibfnamefont {W.}~\bibnamefont {Barthlott}}\ and\ \bibinfo {author} {\bibfnamefont {C.}~\bibnamefont {Neinhuis}},\ }\bibfield  {title} {\bibinfo {title} {Purity of the sacred lotus, or escape from contamination in biological surfaces},\ }\href {https://doi.org/10.1007/s004250050096} {\bibfield  {journal} {\bibinfo  {journal} {Planta}\ }\textbf {\bibinfo {volume} {202}},\ \bibinfo {pages} {1} (\bibinfo {year} {1997})}\BibitemShut {NoStop}%
\bibitem [{\citenamefont {Lenz}\ \emph {et~al.}(2021)\citenamefont {Lenz}, \citenamefont {Bauer},\ and\ \citenamefont {Ruxton}}]{lenz_ecological_2021}%
  \BibitemOpen
  \bibfield  {author} {\bibinfo {author} {\bibfnamefont {A.-K.}\ \bibnamefont {Lenz}}, \bibinfo {author} {\bibfnamefont {U.}~\bibnamefont {Bauer}},\ and\ \bibinfo {author} {\bibfnamefont {G.~D.}\ \bibnamefont {Ruxton}},\ }\bibfield  {title} {\bibinfo {title} {An ecological perspective on water shedding from leaves},\ }\href {https://doi.org/https://doi.org/10.1093/jxb/erab479} {\bibfield  {journal} {\bibinfo  {journal} {Journal of Experimental Botany}\ }\textbf {\bibinfo {volume} {73}},\ \bibinfo {pages} {1176} (\bibinfo {year} {2021})}\BibitemShut {NoStop}%
\bibitem [{\citenamefont {Rose}(2002)}]{rose_dropwise_2002}%
  \BibitemOpen
  \bibfield  {author} {\bibinfo {author} {\bibfnamefont {J.~W.}\ \bibnamefont {Rose}},\ }\bibfield  {title} {\bibinfo {title} {Dropwise condensation theory and experiment: {A} review},\ }\href {https://doi.org/10.1243/09576500260049034} {\bibfield  {journal} {\bibinfo  {journal} {Proceedings of the Institution of Mechanical Engineers, Part A: Journal of Power and Energy}\ }\textbf {\bibinfo {volume} {216}},\ \bibinfo {pages} {115} (\bibinfo {year} {2002})}\BibitemShut {NoStop}%
\bibitem [{\citenamefont {Kim}\ and\ \citenamefont {Kim}(2011)}]{kim_dropwise_2011}%
  \BibitemOpen
  \bibfield  {author} {\bibinfo {author} {\bibfnamefont {S.}~\bibnamefont {Kim}}\ and\ \bibinfo {author} {\bibfnamefont {K.~J.}\ \bibnamefont {Kim}},\ }\bibfield  {title} {\bibinfo {title} {Dropwise condensation modeling suitable for superhydrophobic surfaces},\ }\href {https://doi.org/10.1115/1.4003742} {\bibfield  {journal} {\bibinfo  {journal} {J. Heat Transfer}\ }\textbf {\bibinfo {volume} {133}},\ \bibinfo {pages} {081502} (\bibinfo {year} {2011})}\BibitemShut {NoStop}%
\bibitem [{\citenamefont {Narhe}\ and\ \citenamefont {Beysens}(2004)}]{narhe_nucleation_2004}%
  \BibitemOpen
  \bibfield  {author} {\bibinfo {author} {\bibfnamefont {R.~D.}\ \bibnamefont {Narhe}}\ and\ \bibinfo {author} {\bibfnamefont {D.~A.}\ \bibnamefont {Beysens}},\ }\bibfield  {title} {\bibinfo {title} {Nucleation and {Growth} on a {Superhydrophobic} {Grooved} {Surface}},\ }\bibfield  {journal} {\bibinfo  {journal} {Physical Review Letters}\ }\textbf {\bibinfo {volume} {93}},\ \href {https://doi.org/10.1103/PhysRevLett.93.076103} {10.1103/PhysRevLett.93.076103} (\bibinfo {year} {2004})\BibitemShut {NoStop}%
\bibitem [{\citenamefont {Wier}\ and\ \citenamefont {McCarthy}(2006)}]{wier_condensation_2006}%
  \BibitemOpen
  \bibfield  {author} {\bibinfo {author} {\bibfnamefont {K.}~\bibnamefont {Wier}}\ and\ \bibinfo {author} {\bibfnamefont {T.}~\bibnamefont {McCarthy}},\ }\bibfield  {title} {\bibinfo {title} {Condensation on {Ultrahydrophobic} {Surfaces} and {Its} {Effect} on {Droplet} {Mobility}: {Ultrahydrophobic} {Surfaces} {Are} {Not} {Always} {Water} {Repellant}},\ }\href {https://doi.org/10.1021/la0525877} {\bibfield  {journal} {\bibinfo  {journal} {Langmuir}\ }\textbf {\bibinfo {volume} {22}},\ \bibinfo {pages} {2433} (\bibinfo {year} {2006})}\BibitemShut {NoStop}%
\bibitem [{\citenamefont {{UN Water}}(2020)}]{un_water_2020}%
  \BibitemOpen
  \bibinfo {editor} {\bibnamefont {{UN Water}}},\ ed.,\ \href {https://unesdoc.unesco.org/ark:/48223/pf0000372985.locale=en} {\emph {\bibinfo {title} {Water and climate change}}},\ \bibinfo {series} {The {United} {Nations} world water development report}\ No.\ \bibinfo {number} {2020}\ (\bibinfo  {publisher} {UNESCO},\ \bibinfo {address} {Paris},\ \bibinfo {year} {2020})\BibitemShut {NoStop}%
\bibitem [{\citenamefont {Shi}\ \emph {et~al.}(2018)\citenamefont {Shi}, \citenamefont {Anderson}, \citenamefont {Tulkoff}, \citenamefont {Kennedy},\ and\ \citenamefont {Boreyko}}]{shi_fog_2018}%
  \BibitemOpen
  \bibfield  {author} {\bibinfo {author} {\bibfnamefont {W.}~\bibnamefont {Shi}}, \bibinfo {author} {\bibfnamefont {M.~J.}\ \bibnamefont {Anderson}}, \bibinfo {author} {\bibfnamefont {J.~B.}\ \bibnamefont {Tulkoff}}, \bibinfo {author} {\bibfnamefont {B.~S.}\ \bibnamefont {Kennedy}},\ and\ \bibinfo {author} {\bibfnamefont {J.~B.}\ \bibnamefont {Boreyko}},\ }\bibfield  {title} {\bibinfo {title} {Fog {Harvesting} with {Harps}},\ }\href {https://doi.org/10.1021/acsami.7b17488} {\bibfield  {journal} {\bibinfo  {journal} {ACS Applied Materials \& Interfaces}\ }\textbf {\bibinfo {volume} {10}},\ \bibinfo {pages} {11979} (\bibinfo {year} {2018})}\BibitemShut {NoStop}%
\bibitem [{\citenamefont {Muselli}\ \emph {et~al.}(2009)\citenamefont {Muselli}, \citenamefont {Beysens}, \citenamefont {Mileta},\ and\ \citenamefont {Milimouk}}]{muselli_dew_2009}%
  \BibitemOpen
  \bibfield  {author} {\bibinfo {author} {\bibfnamefont {M.}~\bibnamefont {Muselli}}, \bibinfo {author} {\bibfnamefont {D.}~\bibnamefont {Beysens}}, \bibinfo {author} {\bibfnamefont {M.}~\bibnamefont {Mileta}},\ and\ \bibinfo {author} {\bibfnamefont {I.}~\bibnamefont {Milimouk}},\ }\bibfield  {title} {\bibinfo {title} {Dew and rain water collection in the {Dalmatian} {Coast}, {Croatia}},\ }\href {https://doi.org/10.1016/j.atmosres.2009.01.004} {\bibfield  {journal} {\bibinfo  {journal} {Atmospheric Research}\ }\textbf {\bibinfo {volume} {92}},\ \bibinfo {pages} {455} (\bibinfo {year} {2009})}\BibitemShut {NoStop}%
\bibitem [{\citenamefont {Van~Hulle}\ and\ \citenamefont {Vandewalle}(2023)}]{van_hulle_effect_2023}%
  \BibitemOpen
  \bibfield  {author} {\bibinfo {author} {\bibfnamefont {J.}~\bibnamefont {Van~Hulle}}\ and\ \bibinfo {author} {\bibfnamefont {N.}~\bibnamefont {Vandewalle}},\ }\bibfield  {title} {\bibinfo {title} {Effect of groove curvature on droplet spreading},\ }\href {https://doi.org/10.1039/D3SM00715D} {\bibfield  {journal} {\bibinfo  {journal} {Soft Matter}\ }\textbf {\bibinfo {volume} {19}},\ \bibinfo {pages} {4669} (\bibinfo {year} {2023})}\BibitemShut {NoStop}%
\bibitem [{\citenamefont {Chen}\ \emph {et~al.}(2018)\citenamefont {Chen}, \citenamefont {Ran}, \citenamefont {Gan}, \citenamefont {Zhou}, \citenamefont {Zhang}, \citenamefont {Zhang}, \citenamefont {Zhang},\ and\ \citenamefont {Jiang}}]{chen_ultrafast_2018}%
  \BibitemOpen
  \bibfield  {author} {\bibinfo {author} {\bibfnamefont {H.}~\bibnamefont {Chen}}, \bibinfo {author} {\bibfnamefont {T.}~\bibnamefont {Ran}}, \bibinfo {author} {\bibfnamefont {Y.}~\bibnamefont {Gan}}, \bibinfo {author} {\bibfnamefont {J.}~\bibnamefont {Zhou}}, \bibinfo {author} {\bibfnamefont {Y.}~\bibnamefont {Zhang}}, \bibinfo {author} {\bibfnamefont {L.}~\bibnamefont {Zhang}}, \bibinfo {author} {\bibfnamefont {D.}~\bibnamefont {Zhang}},\ and\ \bibinfo {author} {\bibfnamefont {L.}~\bibnamefont {Jiang}},\ }\bibfield  {title} {\bibinfo {title} {Ultrafast water harvesting and transport in hierarchical microchannels},\ }\href {https://doi.org/10.1038/s41563-018-0171-9} {\bibfield  {journal} {\bibinfo  {journal} {Nature Materials}\ }\textbf {\bibinfo {volume} {17}},\ \bibinfo {pages} {935} (\bibinfo {year} {2018})}\BibitemShut {NoStop}%
\bibitem [{\citenamefont {Leonard}\ \emph {et~al.}(2023)\citenamefont {Leonard}, \citenamefont {Van~Hulle}, \citenamefont {Weyer}, \citenamefont {Terwagne},\ and\ \citenamefont {Vandewalle}}]{leonard_droplets_2023}%
  \BibitemOpen
  \bibfield  {author} {\bibinfo {author} {\bibfnamefont {M.}~\bibnamefont {Leonard}}, \bibinfo {author} {\bibfnamefont {J.}~\bibnamefont {Van~Hulle}}, \bibinfo {author} {\bibfnamefont {F.}~\bibnamefont {Weyer}}, \bibinfo {author} {\bibfnamefont {D.}~\bibnamefont {Terwagne}},\ and\ \bibinfo {author} {\bibfnamefont {N.}~\bibnamefont {Vandewalle}},\ }\bibfield  {title} {\bibinfo {title} {Droplets sliding on single and multiple vertical fibers},\ }\href {https://doi.org/10.1103/PhysRevFluids.8.103601} {\bibfield  {journal} {\bibinfo  {journal} {Physical Review Fluids}\ }\textbf {\bibinfo {volume} {8}},\ \bibinfo {pages} {103601} (\bibinfo {year} {2023})}\BibitemShut {NoStop}%
\bibitem [{\citenamefont {Van~Hulle}\ \emph {et~al.}(2021)\citenamefont {Van~Hulle}, \citenamefont {Weyer}, \citenamefont {Dorbolo},\ and\ \citenamefont {Vandewalle}}]{van_hulle_capillary_2021}%
  \BibitemOpen
  \bibfield  {author} {\bibinfo {author} {\bibfnamefont {J.}~\bibnamefont {Van~Hulle}}, \bibinfo {author} {\bibfnamefont {F.}~\bibnamefont {Weyer}}, \bibinfo {author} {\bibfnamefont {S.}~\bibnamefont {Dorbolo}},\ and\ \bibinfo {author} {\bibfnamefont {N.}~\bibnamefont {Vandewalle}},\ }\bibfield  {title} {\bibinfo {title} {Capillary transport from barrel to clamshell droplets on conical fibers},\ }\href {https://doi.org/10.1103/PhysRevFluids.6.024501} {\bibfield  {journal} {\bibinfo  {journal} {Physical Review Fluids}\ }\textbf {\bibinfo {volume} {6}},\ \bibinfo {pages} {024501} (\bibinfo {year} {2021})}\BibitemShut {NoStop}%
\bibitem [{\citenamefont {Andrews}\ \emph {et~al.}(2011)\citenamefont {Andrews}, \citenamefont {Eccles}, \citenamefont {Schofield},\ and\ \citenamefont {Badyal}}]{andrews_three-dimensional_2011}%
  \BibitemOpen
  \bibfield  {author} {\bibinfo {author} {\bibfnamefont {H.~G.}\ \bibnamefont {Andrews}}, \bibinfo {author} {\bibfnamefont {E.~A.}\ \bibnamefont {Eccles}}, \bibinfo {author} {\bibfnamefont {W.~C.~E.}\ \bibnamefont {Schofield}},\ and\ \bibinfo {author} {\bibfnamefont {J.~P.~S.}\ \bibnamefont {Badyal}},\ }\bibfield  {title} {\bibinfo {title} {Three-{Dimensional} {Hierarchical} {Structures} for {Fog} {Harvesting}},\ }\href {https://doi.org/10.1021/la2000014} {\bibfield  {journal} {\bibinfo  {journal} {Langmuir}\ }\textbf {\bibinfo {volume} {27}},\ \bibinfo {pages} {3798} (\bibinfo {year} {2011})}\BibitemShut {NoStop}%
\bibitem [{\citenamefont {Protiere}\ \emph {et~al.}(2013)\citenamefont {Protiere}, \citenamefont {Duprat},\ and\ \citenamefont {Stone}}]{protiere_wetting_2013}%
  \BibitemOpen
  \bibfield  {author} {\bibinfo {author} {\bibfnamefont {S.}~\bibnamefont {Protiere}}, \bibinfo {author} {\bibfnamefont {C.}~\bibnamefont {Duprat}},\ and\ \bibinfo {author} {\bibfnamefont {H.~A.}\ \bibnamefont {Stone}},\ }\bibfield  {title} {\bibinfo {title} {Wetting on two parallel fibers: drop to column transitions},\ }\href {https://doi.org/10.1039/C2SM27075G} {\bibfield  {journal} {\bibinfo  {journal} {Soft Matter}\ }\textbf {\bibinfo {volume} {9}},\ \bibinfo {pages} {271} (\bibinfo {year} {2013})}\BibitemShut {NoStop}%
\bibitem [{\citenamefont {Park}\ \emph {et~al.}(2016)\citenamefont {Park}, \citenamefont {Kim}, \citenamefont {Grinthal}, \citenamefont {He}, \citenamefont {Fox}, \citenamefont {Weaver},\ and\ \citenamefont {Aizenberg}}]{park_condensation_2016}%
  \BibitemOpen
  \bibfield  {author} {\bibinfo {author} {\bibfnamefont {K.-C.}\ \bibnamefont {Park}}, \bibinfo {author} {\bibfnamefont {P.}~\bibnamefont {Kim}}, \bibinfo {author} {\bibfnamefont {A.}~\bibnamefont {Grinthal}}, \bibinfo {author} {\bibfnamefont {N.}~\bibnamefont {He}}, \bibinfo {author} {\bibfnamefont {D.}~\bibnamefont {Fox}}, \bibinfo {author} {\bibfnamefont {J.~C.}\ \bibnamefont {Weaver}},\ and\ \bibinfo {author} {\bibfnamefont {J.}~\bibnamefont {Aizenberg}},\ }\bibfield  {title} {\bibinfo {title} {Condensation on slippery asymmetric bumps},\ }\href {https://doi.org/10.1038/nature16956} {\bibfield  {journal} {\bibinfo  {journal} {Nature}\ }\textbf {\bibinfo {volume} {531}},\ \bibinfo {pages} {78} (\bibinfo {year} {2016})}\BibitemShut {NoStop}%
\bibitem [{\citenamefont {Bintein}\ \emph {et~al.}(2019)\citenamefont {Bintein}, \citenamefont {Lhuissier}, \citenamefont {Mongruel}, \citenamefont {Royon},\ and\ \citenamefont {Beysens}}]{bintein_grooves_2019}%
  \BibitemOpen
  \bibfield  {author} {\bibinfo {author} {\bibfnamefont {P.-B.}\ \bibnamefont {Bintein}}, \bibinfo {author} {\bibfnamefont {H.}~\bibnamefont {Lhuissier}}, \bibinfo {author} {\bibfnamefont {A.}~\bibnamefont {Mongruel}}, \bibinfo {author} {\bibfnamefont {L.}~\bibnamefont {Royon}},\ and\ \bibinfo {author} {\bibfnamefont {D.}~\bibnamefont {Beysens}},\ }\bibfield  {title} {\bibinfo {title} {Grooves {Accelerate} {Dew} {Shedding}},\ }\href {https://doi.org/10.1103/PhysRevLett.122.098005} {\bibfield  {journal} {\bibinfo  {journal} {Physical Review Letters}\ }\textbf {\bibinfo {volume} {122}},\ \bibinfo {pages} {098005} (\bibinfo {year} {2019})}\BibitemShut {NoStop}%
\bibitem [{\citenamefont {Lavielle}\ \emph {et~al.}(2023{\natexlab{a}})\citenamefont {Lavielle}, \citenamefont {Mongruel}, \citenamefont {Bourouina}, \citenamefont {Royon},\ and\ \citenamefont {Beysens}}]{lavielle_plastic_2023}%
  \BibitemOpen
  \bibfield  {author} {\bibinfo {author} {\bibfnamefont {N.}~\bibnamefont {Lavielle}}, \bibinfo {author} {\bibfnamefont {A.}~\bibnamefont {Mongruel}}, \bibinfo {author} {\bibfnamefont {T.}~\bibnamefont {Bourouina}}, \bibinfo {author} {\bibfnamefont {L.}~\bibnamefont {Royon}},\ and\ \bibinfo {author} {\bibfnamefont {D.}~\bibnamefont {Beysens}},\ }\bibfield  {title} {\bibinfo {title} {Plastic foil micro-grooved by embossing enhances dew collection without aging effects},\ }\href {https://doi.org/10.1016/j.mtsust.2023.100566} {\bibfield  {journal} {\bibinfo  {journal} {Materials Today Sustainability}\ }\textbf {\bibinfo {volume} {24}},\ \bibinfo {pages} {100566} (\bibinfo {year} {2023}{\natexlab{a}})}\BibitemShut {NoStop}%
\bibitem [{\citenamefont {Pou‐Álvarez}\ \emph {et~al.}(2025)\citenamefont {Pou‐Álvarez}, \citenamefont {Mongruel}, \citenamefont {Lavielle}, \citenamefont {Riveiro}, \citenamefont {Bourouina}, \citenamefont {Royon}, \citenamefont {Pou},\ and\ \citenamefont {Beysens}}]{poualvarez_efficient_2025}%
  \BibitemOpen
  \bibfield  {author} {\bibinfo {author} {\bibfnamefont {P.}~\bibnamefont {Pou‐Álvarez}}, \bibinfo {author} {\bibfnamefont {A.}~\bibnamefont {Mongruel}}, \bibinfo {author} {\bibfnamefont {N.}~\bibnamefont {Lavielle}}, \bibinfo {author} {\bibfnamefont {A.}~\bibnamefont {Riveiro}}, \bibinfo {author} {\bibfnamefont {T.}~\bibnamefont {Bourouina}}, \bibinfo {author} {\bibfnamefont {L.}~\bibnamefont {Royon}}, \bibinfo {author} {\bibfnamefont {J.}~\bibnamefont {Pou}},\ and\ \bibinfo {author} {\bibfnamefont {D.}~\bibnamefont {Beysens}},\ }\bibfield  {title} {\bibinfo {title} {Efficient {Autonomous} {Dew} {Water} {Harvesting} by {Laser} {Micropatterning}: {Superhydrophilic} and {High} {Emissivity} {Robust} {Grooved} {Metallic} {Surfaces} {Enabling} {Filmwise} {Condensation} and {Radiative} {Cooling}},\ }\href {https://doi.org/10.1002/adma.202419472} {\bibfield  {journal} {\bibinfo  {journal} {Advanced Materials}\ ,\ \bibinfo {pages} {2419472}} (\bibinfo {year} {2025})}\BibitemShut {NoStop}%
\bibitem [{\citenamefont {Trosseille}\ \emph {et~al.}(2019)\citenamefont {Trosseille}, \citenamefont {Mongruel}, \citenamefont {Royon}, \citenamefont {Medici},\ and\ \citenamefont {Beysens}}]{trosseille_roughness-enhanced_2019}%
  \BibitemOpen
  \bibfield  {author} {\bibinfo {author} {\bibfnamefont {J.}~\bibnamefont {Trosseille}}, \bibinfo {author} {\bibfnamefont {A.}~\bibnamefont {Mongruel}}, \bibinfo {author} {\bibfnamefont {L.}~\bibnamefont {Royon}}, \bibinfo {author} {\bibfnamefont {M.-G.}\ \bibnamefont {Medici}},\ and\ \bibinfo {author} {\bibfnamefont {D.}~\bibnamefont {Beysens}},\ }\bibfield  {title} {\bibinfo {title} {Roughness-enhanced collection of condensed droplets},\ }\href {https://doi.org/10.1140/epje/i2019-11905-9} {\bibfield  {journal} {\bibinfo  {journal} {The European Physical Journal E}\ }\textbf {\bibinfo {volume} {42}},\ \bibinfo {pages} {144} (\bibinfo {year} {2019})}\BibitemShut {NoStop}%
\bibitem [{\citenamefont {Jin}\ \emph {et~al.}(2017)\citenamefont {Jin}, \citenamefont {Zhang},\ and\ \citenamefont {Wang}}]{jin_atmospheric_2017}%
  \BibitemOpen
  \bibfield  {author} {\bibinfo {author} {\bibfnamefont {Y.}~\bibnamefont {Jin}}, \bibinfo {author} {\bibfnamefont {L.}~\bibnamefont {Zhang}},\ and\ \bibinfo {author} {\bibfnamefont {P.}~\bibnamefont {Wang}},\ }\bibfield  {title} {\bibinfo {title} {Atmospheric {Water} {Harvesting}: {Role} of {Surface} {Wettability} and {Edge} {Effect}},\ }\href {https://doi.org/10.1002/gch2.201700019} {\bibfield  {journal} {\bibinfo  {journal} {Global Challenges}\ }\textbf {\bibinfo {volume} {1}},\ \bibinfo {pages} {1700019} (\bibinfo {year} {2017})}\BibitemShut {NoStop}%
\bibitem [{\citenamefont {Jacobs}\ \emph {et~al.}(2008)\citenamefont {Jacobs}, \citenamefont {Heusinkveld},\ and\ \citenamefont {Berkowicz}}]{jacobs_passive_2008}%
  \BibitemOpen
  \bibfield  {author} {\bibinfo {author} {\bibfnamefont {A.}~\bibnamefont {Jacobs}}, \bibinfo {author} {\bibfnamefont {B.}~\bibnamefont {Heusinkveld}},\ and\ \bibinfo {author} {\bibfnamefont {S.}~\bibnamefont {Berkowicz}},\ }\bibfield  {title} {\bibinfo {title} {Passive dew collection in a grassland area, {The} {Netherlands}},\ }\href {https://doi.org/10.1016/j.atmosres.2007.06.007} {\bibfield  {journal} {\bibinfo  {journal} {Atmospheric Research}\ }\textbf {\bibinfo {volume} {87}},\ \bibinfo {pages} {377} (\bibinfo {year} {2008})}\BibitemShut {NoStop}%
\bibitem [{\citenamefont {Gittens}(1969)}]{gittens_variation_1969}%
  \BibitemOpen
  \bibfield  {author} {\bibinfo {author} {\bibfnamefont {G.}~\bibnamefont {Gittens}},\ }\bibfield  {title} {\bibinfo {title} {Variation of surface tension of water with temperature},\ }\href {https://doi.org/10.1016/0021-9797(69)90409-3} {\bibfield  {journal} {\bibinfo  {journal} {Journal of Colloid and Interface Science}\ }\textbf {\bibinfo {volume} {30}},\ \bibinfo {pages} {406} (\bibinfo {year} {1969})}\BibitemShut {NoStop}%
\bibitem [{\citenamefont {Lavielle}\ \emph {et~al.}(2023{\natexlab{b}})\citenamefont {Lavielle}, \citenamefont {Beysens},\ and\ \citenamefont {Mongruel}}]{lavielle_memory_2023}%
  \BibitemOpen
  \bibfield  {author} {\bibinfo {author} {\bibfnamefont {N.}~\bibnamefont {Lavielle}}, \bibinfo {author} {\bibfnamefont {D.}~\bibnamefont {Beysens}},\ and\ \bibinfo {author} {\bibfnamefont {A.}~\bibnamefont {Mongruel}},\ }\bibfield  {title} {\bibinfo {title} {Memory {Re}-{Condensation}},\ }\href {https://doi.org/10.1021/acs.langmuir.2c03070} {\bibfield  {journal} {\bibinfo  {journal} {Langmuir}\ }\textbf {\bibinfo {volume} {39}},\ \bibinfo {pages} {2008} (\bibinfo {year} {2023}{\natexlab{b}})}\BibitemShut {NoStop}%
\bibitem [{\citenamefont {Extrand}\ and\ \citenamefont {Gent}(1990)}]{extrand_retention_1990}%
  \BibitemOpen
  \bibfield  {author} {\bibinfo {author} {\bibfnamefont {C.}~\bibnamefont {Extrand}}\ and\ \bibinfo {author} {\bibfnamefont {A.}~\bibnamefont {Gent}},\ }\bibfield  {title} {\bibinfo {title} {Retention of liquid drops by solid surfaces},\ }\href {https://doi.org/10.1016/0021-9797(90)90225-D} {\bibfield  {journal} {\bibinfo  {journal} {Journal of Colloid and Interface Science}\ }\textbf {\bibinfo {volume} {138}},\ \bibinfo {pages} {431} (\bibinfo {year} {1990})}\BibitemShut {NoStop}%
\bibitem [{\citenamefont {Gao}\ \emph {et~al.}(2018)\citenamefont {Gao}, \citenamefont {Geyer}, \citenamefont {Pilat}, \citenamefont {Wooh}, \citenamefont {Vollmer}, \citenamefont {Butt},\ and\ \citenamefont {Berger}}]{gao_how_2018}%
  \BibitemOpen
  \bibfield  {author} {\bibinfo {author} {\bibfnamefont {N.}~\bibnamefont {Gao}}, \bibinfo {author} {\bibfnamefont {F.}~\bibnamefont {Geyer}}, \bibinfo {author} {\bibfnamefont {D.~W.}\ \bibnamefont {Pilat}}, \bibinfo {author} {\bibfnamefont {S.}~\bibnamefont {Wooh}}, \bibinfo {author} {\bibfnamefont {D.}~\bibnamefont {Vollmer}}, \bibinfo {author} {\bibfnamefont {H.-J.}\ \bibnamefont {Butt}},\ and\ \bibinfo {author} {\bibfnamefont {R.}~\bibnamefont {Berger}},\ }\bibfield  {title} {\bibinfo {title} {How drops start sliding over solid surfaces},\ }\href {https://doi.org/10.1038/nphys4305} {\bibfield  {journal} {\bibinfo  {journal} {Nature Physics}\ }\textbf {\bibinfo {volume} {14}},\ \bibinfo {pages} {191} (\bibinfo {year} {2018})}\BibitemShut {NoStop}%
\bibitem [{\citenamefont {Podgorski}\ \emph {et~al.}(2001)\citenamefont {Podgorski}, \citenamefont {Flesselles},\ and\ \citenamefont {Limat}}]{podgorski_corners_2001}%
  \BibitemOpen
  \bibfield  {author} {\bibinfo {author} {\bibfnamefont {T.}~\bibnamefont {Podgorski}}, \bibinfo {author} {\bibfnamefont {J.-M.}\ \bibnamefont {Flesselles}},\ and\ \bibinfo {author} {\bibfnamefont {L.}~\bibnamefont {Limat}},\ }\bibfield  {title} {\bibinfo {title} {Corners, {Cusps}, and {Pearls} in {Running} {Drops}},\ }\href {https://doi.org/10.1103/PhysRevLett.87.036102} {\bibfield  {journal} {\bibinfo  {journal} {Physical Review Letters}\ }\textbf {\bibinfo {volume} {87}},\ \bibinfo {pages} {036102} (\bibinfo {year} {2001})}\BibitemShut {NoStop}%
\bibitem [{\citenamefont {Le~Grand}\ \emph {et~al.}(2005)\citenamefont {Le~Grand}, \citenamefont {Daerr},\ and\ \citenamefont {Limat}}]{le_grand_shape_2005}%
  \BibitemOpen
  \bibfield  {author} {\bibinfo {author} {\bibfnamefont {N.}~\bibnamefont {Le~Grand}}, \bibinfo {author} {\bibfnamefont {A.}~\bibnamefont {Daerr}},\ and\ \bibinfo {author} {\bibfnamefont {L.}~\bibnamefont {Limat}},\ }\bibfield  {title} {\bibinfo {title} {Shape and motion of drops sliding down an inclined plane},\ }\href {https://doi.org/10.1017/S0022112005006105} {\bibfield  {journal} {\bibinfo  {journal} {Journal of Fluid Mechanics}\ }\textbf {\bibinfo {volume} {541}},\ \bibinfo {pages} {293} (\bibinfo {year} {2005})}\BibitemShut {NoStop}%
\bibitem [{\citenamefont {Seemann}\ \emph {et~al.}(2005)\citenamefont {Seemann}, \citenamefont {Brinkmann}, \citenamefont {Kramer}, \citenamefont {Lange},\ and\ \citenamefont {Lipowsky}}]{seemann_wetting_2005}%
  \BibitemOpen
  \bibfield  {author} {\bibinfo {author} {\bibfnamefont {R.}~\bibnamefont {Seemann}}, \bibinfo {author} {\bibfnamefont {M.}~\bibnamefont {Brinkmann}}, \bibinfo {author} {\bibfnamefont {E.~J.}\ \bibnamefont {Kramer}}, \bibinfo {author} {\bibfnamefont {F.~F.}\ \bibnamefont {Lange}},\ and\ \bibinfo {author} {\bibfnamefont {R.}~\bibnamefont {Lipowsky}},\ }\bibfield  {title} {\bibinfo {title} {Wetting morphologies at microstructured surfaces},\ }\href {https://doi.org/10.1073/pnas.0407721102} {\bibfield  {journal} {\bibinfo  {journal} {Proceedings of the National Academy of Sciences}\ }\textbf {\bibinfo {volume} {102}},\ \bibinfo {pages} {1848} (\bibinfo {year} {2005})}\BibitemShut {NoStop}%
\bibitem [{\citenamefont {Seemann}\ \emph {et~al.}(2011)\citenamefont {Seemann}, \citenamefont {Brinkmann}, \citenamefont {Herminghaus}, \citenamefont {Khare}, \citenamefont {Law}, \citenamefont {McBride}, \citenamefont {Kostourou}, \citenamefont {Gurevich}, \citenamefont {Bommer}, \citenamefont {Herrmann},\ and\ \citenamefont {Michler}}]{seemann_wetting_2011}%
  \BibitemOpen
  \bibfield  {author} {\bibinfo {author} {\bibfnamefont {R.}~\bibnamefont {Seemann}}, \bibinfo {author} {\bibfnamefont {M.}~\bibnamefont {Brinkmann}}, \bibinfo {author} {\bibfnamefont {S.}~\bibnamefont {Herminghaus}}, \bibinfo {author} {\bibfnamefont {K.}~\bibnamefont {Khare}}, \bibinfo {author} {\bibfnamefont {B.~M.}\ \bibnamefont {Law}}, \bibinfo {author} {\bibfnamefont {S.}~\bibnamefont {McBride}}, \bibinfo {author} {\bibfnamefont {K.}~\bibnamefont {Kostourou}}, \bibinfo {author} {\bibfnamefont {E.}~\bibnamefont {Gurevich}}, \bibinfo {author} {\bibfnamefont {S.}~\bibnamefont {Bommer}}, \bibinfo {author} {\bibfnamefont {C.}~\bibnamefont {Herrmann}},\ and\ \bibinfo {author} {\bibfnamefont {D.}~\bibnamefont {Michler}},\ }\bibfield  {title} {\bibinfo {title} {Wetting morphologies and their transitions in grooved substrates},\ }\href {https://doi.org/10.1088/0953-8984/23/18/184108} {\bibfield  {journal} {\bibinfo  {journal} {Journal of Physics: Condensed Matter}\ }\textbf {\bibinfo {volume} {23}},\ \bibinfo
  {pages} {184108} (\bibinfo {year} {2011})}\BibitemShut {NoStop}%
\bibitem [{\citenamefont {Narhe}\ and\ \citenamefont {Beysens}(2006)}]{narhe_water_2006}%
  \BibitemOpen
  \bibfield  {author} {\bibinfo {author} {\bibfnamefont {R.~D.}\ \bibnamefont {Narhe}}\ and\ \bibinfo {author} {\bibfnamefont {D.~A.}\ \bibnamefont {Beysens}},\ }\bibfield  {title} {\bibinfo {title} {Water condensation on a super-hydrophobic spike surface},\ }\href {https://doi.org/10.1209/epl/i2006-10069-9} {\bibfield  {journal} {\bibinfo  {journal} {Europhysics Letters (EPL)}\ }\textbf {\bibinfo {volume} {75}},\ \bibinfo {pages} {98} (\bibinfo {year} {2006})}\BibitemShut {NoStop}%
\bibitem [{\citenamefont {Beysens}(2018)}]{beysens_dew_2018}%
  \BibitemOpen
  \bibfield  {author} {\bibinfo {author} {\bibfnamefont {D.}~\bibnamefont {Beysens}},\ }\href {https://doi.org/10.1201/9781003337898} {\emph {\bibinfo {title} {Dew {Water}}}},\ \bibinfo {edition} {1st}\ ed.\ (\bibinfo  {publisher} {River Publishers},\ \bibinfo {year} {2018})\BibitemShut {NoStop}%
\end{thebibliography}
\end{document}